\documentclass[aps, prb, 10pt, twocolumn]{revtex4-1}
\usepackage[english]{babel}
\usepackage[breaklinks=true,colorlinks=true,citecolor=blue,urlcolor=blue,bookmarks=false]{hyperref}
\usepackage{amsmath,amssymb,amsfonts}

\usepackage{todonotes}

\newcommand{\ph}{\varphi}

\newcommand{\ov}[1]{\bar{#1}}

\newcommand{\V}{\mathcal{V}}

\newcommand{\tr}[1]{\text{Tr}\bigl[#1\bigr]}

\newcommand{\snu}{\sqrt{\nu}}

\newcommand{\io}{I_{0}}
\newcommand{\hio}{\hat{I}_{0}}

\newcommand{\imax}{I_{\text{max}}}

\newcommand{\htu}{H_{T}}
\newcommand{\ut}{\mathcal{U}}
\newcommand{\dint}[1]{\mathcal{D}t_{#1}}

\newcommand{\ib}{\hat{I}_B^I}

\newcommand{\mo}{\mathcal{O}}

\begin{document}

\title{Non-equilibrium noise in the (non-)Abelian fractional quantum Hall effect}

\author{O. Smits}
\email{smitso@tcd.ie}
\affiliation{Dublin Institute for Advanced Studies, School of Theoretical Physics, 10 Burlington Rd, Dublin, Ireland}
\affiliation{School of Mathematics, Trinity College, Dublin 2, Ireland}
\author{J. K. Slingerland}
\affiliation{Dublin Institute for Advanced Studies, School of Theoretical Physics, 10 Burlington Rd, Dublin, Ireland}
\affiliation{Department of Mathematical Physics, National University of Ireland, Maynooth, Ireland}

\author{S. H. Simon}
\affiliation{Rudolf Peierls Centre for Theoretical Physics, Oxford, OX1 3NP, UK}
\affiliation{Department of Mathematical Physics, National University of Ireland, Maynooth, Ireland}

\date{\today}

\begin{abstract}
We analyse the noise of the edge current of a generic fractional quantum Hall state in a tunnelling point contact system. We show that the non-symmetrized noise in the edge current for the system out-of-equilibrium is completely determined by the noise in the tunnelling current and the Nyquist-Johnson (equilibrium) noise of the edge current. Simply put, the noise in the tunnelling current does not simply add up the equilibrium noise of the edge current. A correction term arises associated with the correlation between the tunnelling current and the edge current. We show, using a non-equilibrium Ward identity, that this correction term is determined by the anti-symmetric part of the noise in the tunnelling current. This leads to a non-equilibrium fluctuation-dissipation theorem and related expressions for the excess and shot noise of the noise in the edge current. Our approach makes use of simple properties of the edge, such as charge conservation and chirality, and applies to generic constructions of the edge theory which includes edges of non-Abelian states and edges with multiple charged channels. Two important tools we make use of are the non-equilibrium Kubo formula and the non-equilibrium Ward identity. We discuss these identities in the appendix.
\end{abstract}

\maketitle

%%%% INTRODUCTION %%%% 

\section{Introduction}
The fractional quantum Hall effect\citep{dassarma1996} is an example of a topological phase of matter\citep{wen2012}. At each plateaux the electrical Hall resistance is quantized and the collective behaviour of the electrons is said to be topologically ordered\citep{wen2004}. Characteristic features of these phases are a topological quantum field theory as the low energy description, the presence of a bulk energy gap, a robustness of the low-energy theory against local perturbations and quasiparticle excitations known as anyons\citep{leinaas1977,laughlin1983,arovas1984}. Non-Abelian anyons in particular obey a very rich generalization of exchange statistics, and the $\nu=5/2$ state has been put forward as a candidate for the realization of these quasiparticles\citep{mooreread1991, greiter1991, nayak1996}. Although much effort has been put into studying this and other candidates phases the experimental discovery of a non-Abelian anyon is as of yet an open question. The stakes are high as non-Abelian anyons could lead to the realization of a topological quantum computer\citep{kitaev2003,dassarma2006,nayak2008}.

The edge of a fractional quantum Hall state is responsible for the transport properties of the system\citep{wen1990a,wen1992}. Edge states are chiral and topologically protected, and backscattering of charge can only occur between opposite edges. Tunnelling experiments in the fractional quantum Hall effect make use of this property and probe the low-energy states of the system through use of a tunnelling point contact\citep{wen1991,kanefisher1992, moon1993}. A tunnelling point contact acts as a constriction which forces opposite edges together and induces tunnelling of (charged) quasiparticles between the edges. This results in a tunnelling current which is characterized by the specific edge theory and the underlying topological order. Because of this both the tunnelling current and its fluctuations (also known as the noise) can be used to identify the topological order of the system. 

\begin{figure}
\includegraphics[width=.4\textwidth]{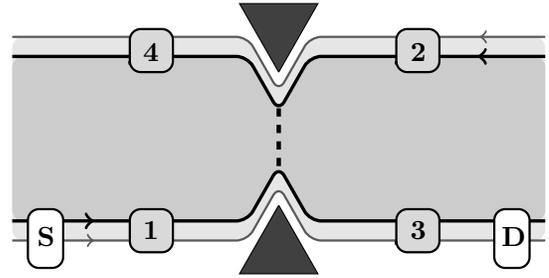}
\caption{Sketch of a point contact. A current is injected at the source (S), flows along the edge and is collected by the drain (D). At the point contact quasiparticles tunnel between the edges and a backscattering current forms flowing from the lower to the upper edge. The probes 1 through 4 can be used to measure the local edge current and the corresponding noise.}\label{fig:pointcontact} 
\end{figure}
The expression for the noise and in particular its relation to the tunnelling current (or, equivalently, the transmission) has been studied perturbatively for general and specific quantum Hall states\citep{kanefisher1994, chamon1995, chamon1996, chang2003, martin2005, bena2006, bena2007,safi2008,ferraro2008,ferraro2010, ferraro2010a, ferraro2010b, carrega2011, iyoda2011, carrega2012}. For special cases such as the integer quantum Hall effect\citep{landauer1991,buttiker1992, martin1992} and the Laughlin series\citep{fendley1995c, fendley1996, trauzettel2004} there are also non-perturbative results. The simplest example of a perturbative approach is the Schottky relation\citep{schottky1926}, which arises in the low temperature and weak tunnelling limit. It relates the shot noise and tunnelling current through $S_{I_B}(0) = e^* I_B$ which can be used to measure the quasiparticle charge. However, a universal expression relating the noise and the current non-perturbatively is still an open question.

Experiments that measure shot noise\citep{saminadayar1997,depicciotto1997,glattli1998,reznikov1999, glattli2000, griffiths2000, heiblum2000, comforti2002, chung2003a, chung2003b, heiblum2003, heiblum2006, dolev2008,chen2009,bid2009, dolev2010,dolev2010b, dolev2011, dolev2011b} do not actually measure the noise in the tunnelling current directly, but instead look at the noise in the outgoing edge currents. To clarify, consider Figure~\ref{fig:pointcontact} which shows a schematic of the experimental setup of a tunnelling point contact. A current is injected at the source (S). It flows along the edge and is partially reflected at the point contact. The dotted line represents the tunnelling current. This current and the corresponding noise are not measured directly, but instead end up in the outgoing branches of the edge currents. A probe located at position 3 or 4 then measures the local edge current and corresponding fluctuations (this probe can also be incorporated with the drain -- here we use a simplistic picture).

This setup then begs the question: how is the noise at, say, probe number 3 related to the noise in the tunnelling current?  In this work we derive such a relation based on general grounds. We use conservation of charge combined with the chiral structure of the edge. Any charge tunnelling from the upper to the lower edge will end up at probe number 3 due to the chiral structure. In this work we study the exact expression relating the noise in the outgoing current to the noise in the tunnelling current. This question has been studied several times before, both non-perturbatively\citep{buttiker1992, kanefisher1994,fendley1995c, bena2007, wang2013} and perturbatively\citep{chamon1996, wang2011, iyoda2011}.

We will give a summary of our approach and our results in the next section. What is important to keep in mind is that the expression which relates the noise in the edge current to the noise in the tunnelling current is \emph{not} linear. The fluctuations of the tunnelling current do not simply add to the fluctuations in the edge  current. The relation between the noise in the edge current and the noise in the tunnelling current is also known as a \emph{non-equilibrium fluctuation-dissipation theorem}.

%%%% SUMMARY %%%% 

\section{Summary and overview of this work}\label{sec:summary}

\begin{figure}
\includegraphics[width=.42\textwidth]{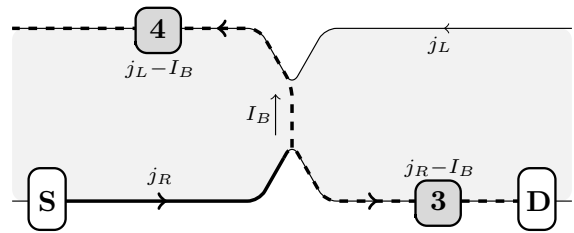}
\caption{Sketch of a point contact. An edge current $j_R$ is injected at the source $S$, and is partially reflected by the point contact resulting in a tunnelling current $I_B$. The edge current after the point contact is therefore $j_R - I_B$ (on the basis of charge conservation and the chiral structure of the edge). The noise measured at probe 3 corresponds to Eq.~\eqref{noise_formula_simple}. In this work this relations are derived at the operator level.}\label{fig:pointcontactsimple}
\end{figure}

Let us present an intuitive picture of the derivation in this work which relates the noise in the outgoing edge current to the noise in the tunnelling current. We start with the simplified Figure~\ref{fig:pointcontactsimple}. A current is injected at $S$ into the lower edge carried by the (right moving) edge current $j_R$. This chiral current is partially reflected by the point contact, where a tunnelling current $I_B$ tunnels to the upper edge and ends up in the left moving edge current. On the basis of charge conservation the edge current that is measured by probe 3 equals
\begin{align}
j_3 = j_R -I_B~.\label{current_equation_simple}
\end{align}
Suppose we now measure the noise of the edge current at probe 3. This noise is roughly given by the square of the operator or
\begin{align}
S_{3} \sim (j_R - I_B)^2 &= j_R^2 + I_B^2 - (j_R I_B + I_Bj_R) \nonumber\\
&\sim S_{\text{bg}} + S_{I_B} - \Delta S~. \label{noise_formula_simple}
\end{align}
In the second line we identify $I_B^2$ with the noise in the tunnelling current $S_{I_B}$ and $S_{\text{bg}}$ a type of background noise (the noise of the edge current in the absence of a point contact). There also appears a third term $\Delta S$, which represents the coupling of the tunnelling current with the equilibrium edge current. This extra term arises because the noise is not linear in the edge current. 

In this work we reproduce this argument \emph{at the operator level}, which is also what distinguishes our approach from previous work~\citep{kanefisher1994, bena2007, wang2011}. We analyse the non-symmetrized noise in the outgoing edge current $S_{3}$ for generic quantum Hall systems. For that we use two new tools, which we have developed in this work. The first is the \emph{non-equilibrium Kubo formula}. This NE-Kubo formula formally extends the expression for linear response theory to all orders of perturbation theory, and from it we obtain the current equation (Kirchoff's law) Eq.~\eqref{current_equation_simple} at the operator level. Using the same logic we obtain a formula of the form Eq.~\eqref{noise_formula_simple}.

The second tool we develop is a \emph{non-equilibrium Ward identity}. A Ward identity is an identity imposed on the correlation function, due to the presence of a symmetry in the theory. In this work the symmetry is associated with charge conservation (and $j_R$ is the associated conserved current), which leads to a well-known Ward identity\citep{knizhnik1984}. We have extended this identity to correlation functions evaluated in the non-equilibrium system. The non-equilibrium Ward identity is used to simplify the expression for the correction term $\Delta S \sim (j_R I_B + I_Bj_R)$. This results in the anti-symmetrized noise of $I_B$, i.e. this correction term is proportional to \mbox{$\Delta S\sim (S_{I_B}(\omega)-S_{I_B}(-\omega))$}.

The final result is an expression for the noise in the edge current related to the noise in the tunnelling current, see Eq.~\eqref{eq:mainresult1}. Therefore to compute the noise in the edge current, we only need to determine the expression for the noise in the tunnelling current which is often easier to obtain and for which more work has been performed. Related to this is an expression for the excess noise Eq.~\eqref{eq:mainresult2}, a non-equilibrium fluctuation-dissipation theorem Eq.~\eqref{noneqfdt} and an expression for the shot-noise limit Eq.~\eqref{eqn:shotnoise}. Our main work focuses on an edge with a single charged channel (described by a chiral boson) and possibly one or multiple neutral channels. In addition we show how the results extend to quantum Hall edges with multiple charged modes, possibly counter propagating. Finally, we also look at similar expressions for the noise at the remaining probes of Figure~\ref{fig:pointcontact}, and the noise of combinations of these probes (i.e. the noise in the voltage difference of probe 3 and 4). All results are valid to all orders of perturbation theory.

An important simplification that we assume is that away from the point contact the edge is described by a collection of free and decoupled channels, each described by a chiral conformal field theory in the long wavelength limit. Interaction effects and disorder, which can for instance cause equilibration of the edge currents after the point contact, are beyond the scope of this work. We also note that this paper looks at the (non-equilibrium) relation between the noise of the edge currents and the tunnelling current. We do not determine the expression for the noise or its relation to the tunnelling current. 

In Section~\ref{sec1:model} we start with a summary of a generic quantum Hall edge. We focus on the definition of the edge current operator in the chiral boson model, the construction of a quasiparticle operator and the non-equal time commutation relations of the theory, and extend this construction to edges with multiple charged channels.

In Section~\ref{sec_tunnelling_pc} we discuss the model of a point contact and in Section~\ref{sec:formalism} we summarize the non-equilibrium formalism. For this formalism we assume that, initially, the point contact is absent and the system is at equilibrium and finite temperature. We also discuss the non-equilibrium Kubo formula, which is proven in Appendix~\ref{app1:kirchoff}. In Section~\ref{sec:edgecurrent} we apply the NE-Kubo formula to the edge current operator which results in an operator-version of Kirchoff's law.

The main results regarding the noise are obtained in Section~\ref{sec:noise}. This makes use of the non-equilibrium Ward identity to simplify the expression for the correction term $\Delta S$. We obtain expressions for the non-symmetrized noise, the excess noise and the shot noise (all of the noise in the edge current) and generalize these expression to the multichannel case. Finally, Section~\ref{app:othernoise} discusses expressions for the noise in related quantities. We discuss our findings in the Section~\ref{sec:conclusion}

\section{The edge of a fractional quantum Hall state}\label{sec1:model}

In this section we discuss the edge theory of a generic fractional quantum Hall state. Before we come to this we emphasize that our main analysis is quite general and does not require all technical details associated with the edge theory. The required input for the treatment on the noise is (1) the edge current operator $j_R$ (2) the quasiparticle operator $\psi^\dag$ which is used to represent quasiparticle tunnelling, and (3) the non-equal time commutation relations of the edge current and quasiparticle operator, Eq.~\eqref{eq:commutation_relation_qp_edge_current}. These relations combined with some basic assumptions, such as translational invariance and chirality, are enough input for our main work which is treated in Section~\ref{sec_tunnelling_pc} and beyond. So although the discussion in this section is somewhat technical and brief, it is only needed to motivate the origin of the edge current operator and to describe the general idea of the edge theory. 

Our treatment of the edge theory is very similar to our previous work in Ref.~\onlinecite{smits2013}. A quantum Hall system is a topologically ordered system, in which chiral gapless states develop at the edge as a consequence of anomaly cancellation\citep{wen1990a,wen1992,bieri2011}. In the long wavelength limit the effective edge theory is a chiral conformal field theory and it comes equipped with a set of quasiparticle operators and fusion rules. Non-Abelian states\citep{mooreread1991,greiter1991, read1999} are characterized by the presence of quasiparticles with multiple fusion channels.

The edge contains a $U(1)$ symmetry due to the coupling with the electromagnetic field. For instance, the edge theory of the Abelian Laughlin state at $\nu = 1/(2m+1)$ is described by a $\hat{u}(1)$ chiral current algebra, also known as the chiral boson or chiral Luttinger liquid\citep{wen1992}. More complicated edge theories are constructed by combining neutral degrees of freedom with one or multiple chiral bosons. These neutral degrees of freedom do not couple to the electromagnetic field and are responsible for the non-Abelian nature of the corresponding trial state. In this work we assume the quasiparticle operators at the edge obey the following decomposition\citep{frohlich2001} 
\begin{align}
\mathcal{A}_{\text{edge}} = \mathcal{W}_n \otimes \hat{u}(1) \otimes \cdots \otimes \hat{u}(1)~. \label{eq:algebrafactorization}
\end{align}
Here the $\hat{u}(1)$'s correspond to the different charged channels of the edge. Since we are interested mostly in the properties of the charged channels we describe all neutral degrees of freedom collectively through $\mathcal{W}_n$. We first discuss the case of a single charged channel, and expand this to the multichannel case at the end of this section.

\subsection{The charged channel in the absence of tunnelling}
We consider the charged channel\citep{wen1990a,levkivskyi2009, vondelft1998,floreanini1987} on the lower edge described by a chiral boson $\ph$ subject to a voltage bias $U$ in the gauge $a_x = 0$. The field is compactified and the action is given by
\begin{align}
S_R &= \frac{1}{4\pi} \int_{\Sigma_R} \left[-\eta_R\partial_t\ph_R\partial_x\ph_R - v_c (\partial_x\ph_R)^2\right]~ dt dx \nonumber\\
&+ \frac{\snu}{2\pi} U_R\int_{\Sigma_R}[\partial_x\ph_R] ~ dt dx  ~.\label{chiral_boson_action}
\end{align}
Throughout this work we mostly focus on a right moving edge current boson, which in the single channel case corresponds to a single right moving chiral boson. It is  coupled to the potential $U_R$ and moves along the edge $\Sigma_R$. We can easily switch to a left moving boson by replacing $R\rightarrow L$. The chirality $\eta_R=1$ is written explicitly (and $\eta_L = -1$). Finally $v_c$ is the edge velocity. Quantization of this action is performed in e.g.~Ref.~\onlinecite{floreanini1987}. The non-local commutation relations are
\begin{align}
[\ph_R(x), \ph_R(x')] &= i\eta_R \pi \text{sgn}(x-x')
\end{align}
with $\text{sgn}(x) = +1$, 0, 1 for the regions \mbox{$x>0$, $x=0$} and \mbox{$x<0$}. Heisenberg's equation of motion results in
\begin{align}
(-\eta_R\partial_t - v_c \partial_x)\ph_R = -\snu  U_R~. \label{eq:eqnofmotion}
\end{align}
Using the equations of motion we can extend the commutation relations to non-equal time
\begin{align}
[\ph_R(x, t), \ph_R(0,0)] &= i\eta_R\pi \text{sgn}(x - \eta_R v_c t) \nonumber\\
[\partial_x\ph_R(x, t), \ph_R(0,0)] &= i\eta_R 2\pi \delta(x - \eta_R v_ct))~. \label{commutationrelations}
\end{align}
The left and right moving bosons commute. The dependency on the combination \mbox{$x\pm v_ct$} reflects the chiral nature of the system and we assume the system is translational invariant. The charge density along the edge is identified with the operator
\begin{align}
\rho_R(x) = \frac{\snu}{2\pi} \partial_x\ph_R(x)~. \label{eq:chargedensity}
\end{align}
The corresponding conserved charge is the electric charge operator
\begin{align}
\mathcal{Q}_R = \frac{\snu}{2\pi}\int_{\Sigma_{R}}  \partial_x\ph_R(x)~dx~.
\end{align}
The total edge Hamiltonian includes the contribution of the neutral channel, which we discuss in Section~\ref{sec:neutral_channel}. Using the electric charge operators, the grand canonical Hamiltonian $K_0$ of the total system is given by
\begin{align}
K_{0} &= H_{0,c} + H_{0,n} - U_R \mathcal{Q}_R  - U_L \mathcal{Q}_L  \label{eq:grandcanonicalHamiltonian}\\
H_{0,c} &= \frac{v_c}{4\pi} \int dx \bigl[(\partial_x\ph_{R})^2 + (\partial_x\ph_{L})^2\bigr]
\end{align}
Here $H_{0,c}$ and $H_{0,n}$ are the Hamiltonians of the charged and neutral channels of both left and right movers. The combination $H_{0,c,R}- U_R \mathcal{Q}_R$ follows from the action Eq.~\eqref{chiral_boson_action}, and we set $H_{0,c} = H_{0,c,L} + H_{0,c,R}$.

Eq.~\eqref{eq:grandcanonicalHamiltonian} is of the form of a grand canonical Hamiltonian \mbox{$K_0 = H_0 - \mu \hat{N}$}, with $\mathcal{Q}$ a generalization of the number operator $\hat{N}$. Although the edges are held at different chemical potentials we still refer to this system and the corresponding Hamiltonian as the equilibrium system. When we include the point contact we refer to the system as out of equilibrium.

\subsection{Edge current operator}
The charge density operator is the zeroth component of a conserved edge current $(\rho(x,t), j(x,t))$. To obtain the edge current density operator we use the continuity equation
\begin{align}
\partial_t\rho_R(x) + \partial_x j_R(x) = 0~. \label{eq:continuityequation}
\end{align}
In terms of the bosonic field the continuity equation reads \mbox{$\partial_x(\frac{\snu}{2\pi} \partial_t\ph(x) + j(x)) = 0$}, which determines the edge current in terms of $\partial_t\ph$ up to an \mbox{$x$-independent} term. This term is set to zero by demanding that the current operator produces the usual Hall relation. The edge current operator is
\begin{align} 
j_R(x) = -\frac{\snu}{2\pi} \partial_t\ph_R(x)~. \label{eq:currentoperator}
\end{align}
Using the equations of motion \eqref{eq:eqnofmotion} we have the alternative form in terms of the charge density operator
\begin{align}
j_R(x) &= \eta_R v_c \rho_R(x) -\eta_R \frac{\nu}{2\pi} U_R
\end{align}
Here we recall that by replacing $R\rightarrow L$ we obtain the left moving chiral boson. The total current running through the system is given by
\begin{align}
\hio(x) = j_R(x) + j_L(x)~. \label{currentoperator}
\end{align}
This total current operator is non-local as it adds the edge current densities on opposite edges. In a more general setting the total current operator is obtained by taking the (bulk + edge) current density operator and integrating along a cross section $\int_{\text{lower edge}}^{\text{upper edge}} J(x,y) dy$, see e.g. Ref.~\onlinecite{kane1995}. This reduces to Eq.~\eqref{currentoperator} when the continuity equation Eq.~\eqref{eq:continuityequation} holds.

We have defined the vacuum such that it is charge neutral. This implies the vanishing of the one-point correlator \mbox{$\langle \rho_R(x)\rangle = \langle\partial_x\ph_{R}(x)\rangle =0$} and we find for the current densities on the edge
\begin{align}
\langle j_R(x,t) \rangle &= -\eta_R\frac{\nu}{2\pi} U_R
\end{align}
The expectation values are with respect to the equilibrium Hamiltonian at finite temperature, i.e. \mbox{$\langle \cdots \rangle = \tr{e^{-\beta K_0} \cdots }$}. For the total current we obtain the familiar Hall relation between voltage and current in the absence of backscattering
\begin{align}
\imax = \langle \hio  \rangle &= \langle j_L\rangle + \langle j_R\rangle = 
\frac{\nu}{2\pi} (U_L - U_R) ~. \label{Hallrelation}
\end{align}
in units where $\hbar = e = 1$. Throughout this work $\imax$ is called the equilibrium current which refers to the current running through the system in the absence of tunnelling between edges. We define $V = U_L - U_R$ as the source-drain voltage.

A tunnelling point contact induces backscattering of charge and this modifies the Hall relation \eqref{Hallrelation}. Concretely a so-called backscattering current flows along the point contact from one edge to the other. On the basis of charge conservation we expect that this modifies the Hall relation to
\begin{align}
I = \imax - \langle I_B\rangle
\end{align}
In this work we will prove this relation on the operator level and we study its effect on the noise in the edge current. 

For later purposes we therefore require the autocorrelator of the current which determines the equilibrium noise of the edge current. We set \mbox{$\Delta j_R(x,t) = j_R(x,t) - \langle j_R(x,t)\rangle$}. Since \mbox{$\langle\partial_x\ph_R \rangle =0$} we have \mbox{$\Delta j_R(x,t) = -v_c\frac{\snu}{2\pi}\partial_x\ph_R$}. This gives for the autocorrelator\citep{vondelft1998}
\begin{align}
S_{j_R}(t) &=  \langle \Delta j_R(0,t) \Delta j_R(0,0)\rangle 
\nonumber\\
&=\frac{\nu}{(2\pi)^2}\frac{(\pi k_BT)^2}{\sin\bigl(\pi k_B T(\delta + it)\bigr)^2}
\end{align}
with $\delta$ a UV regulator. The corresponding Fourier transform is\citep{martin2005}
\begin{align}
S_{j_R}(\omega) =  \omega N(\omega) G~.
\end{align}
where $N(\omega) = \coth(\frac{\omega}{2k_B T}) + 1$ and $G = \frac{\nu}{4\pi}$ is half of the total conductivity of the system (the other half is attributed to the left moving edge).

\subsection{Neutral channel and quasiparticles}\label{sec:neutral_channel}
The neutral channel describes edge degrees of freedom which do not couple to the external voltage bias.  Similar to our previous work\citep{smits2013} we do not specify the exact nature of the neutral part, and only demand that the decomposition \eqref{eq:algebrafactorization} holds. In the case of non-Abelian states it is the neutral channel which is responsible for the non-Abelian nature of the quasiparticle.

In this work we are interested in the properties of the edge current operator. This operator completely decouples from the neutral channel. So although the neutral channel plays an import role in specifying the topological order of the system, it does not explicitly enter the remaining analysis of this work. 
 
With that in mind we now give a short overview of how the neutral channel enters the description of the quasiparticles. The neutral channel is described in the long wavelength limit by some chiral conformal field theory which comes equipped with a consistent set of fusion rules\citep{preskill1998} and some Hamiltonian $H_n$. This Hamiltonian enters the definition of the grand canonical Hamiltonian $K_0$, see Eq.~\eqref{eq:grandcanonicalHamiltonian}. In addition there is some characteristic neutral edge velocity $v_n$, and in general $v_n \neq v_c$. A general quasiparticle operator is of the form
\begin{align}
\psi_R^\dag(x,t) \propto \sigma_R(x, t)~e^{-i\eta_R \frac{Q}{\snu} \ph_R(x,t)} \label{quasiparticlesingle}
\end{align}
The exponential and $\sigma$ operator correspond to the charged and neutral channel, respectively. The neutral channel itself is also chiral (i.e. we have a left- and right moving version $\sigma_{R/L}$), but we will not write this explicitly. The charged operator is normal ordered and we assume the operator is properly normalized, see e.g Ref.~\onlinecite{vondelft1998}. Both operators are characterized by their conformal dimension\citep{difrancesco1995} $h_n$ and $h_c$. In particular, for the charged part we have $h_c = \frac{Q^2}{2\nu}$.

For each quasiparticle operator we also have a conjugate operator which has opposite charge and equal conformal dimension\citep{preskill1998,difrancesco1995}
\begin{align}
\psi_R(x,t) \propto \ov\sigma_R(x, t) e^{i\eta_R\frac{Q}{\snu} \ph_R(x,t)}~.
\end{align}
Here $\ov\sigma$ is the unique operator in the conformal field theory which fuses to the identity with $\sigma$
\begin{align}
\sigma \times \ov\sigma = \mathbf{1} + \ldots~.
\end{align} 
If the right hand side contains multiple fusion products, then the quasiparticle is non-Abelian. In some cases, such as the Moore-Read state\citep{mooreread1991} the neutral part of the quasiparticle operator is self-dual meaning $\sigma = \ov\sigma$.

The quasiparticle operator carries a charge $Q$ measured in units of $e=1$. This follows from the commutation relation with the electric charge operator
\begin{align}
[\mathcal{Q}_R, \psi_R^\dag(x,t)] = Q \psi_R^\dag(x,t)~.
\end{align}
Finally, there is also the commutation relation between the edge current and the quasiparticle operator at non-equal times. Using Eq.~\eqref{commutationrelations} we obtain
\begin{align}
[j_R(x,t), \psi_R^\dag(0,0) ] = 
\eta_R v_c Q \psi_R^\dag(0,0) \delta(x - \eta_R v_ct)~. \label{eq:commutation_relation_qp_edge_current}
\end{align}

\subsection{Generalization to multiple charged channels}\label{sec:multichannel_def}
The single chiral boson model is only sufficient to explain the Laughlin series at filling fraction $\nu = 1/(2M+1)$ with $M$ a positive integer. This construction can be extended through use of neutral channels, which allows for a diverse range of filling fractions. An alternative method is to consider multiple copies of chiral bosons, each of which couples to the electromagnetic field. Both constructions are needed to account for the wide variety of observed filling fractions.

We follow here the treatment of Ref.~\onlinecite{levkivskyi2009} and Ref.~\onlinecite{wen1992}. We assume the bosons are decoupled from each other. The action of the right moving edge is given by
\begin{align}
S_R &= \frac{1}{4\pi} \sum_i \int_{\Sigma_R}  \bigl[-\eta_i \partial_t\ph_i \partial_x \ph_i - v_i (\partial_x\ph_i)^2\bigr]~dt dx \nonumber\\
&+ \frac{1}{4\pi} U_R \sum_i \kappa_i \int \partial_x \ph_i~ dt dx~. \label{multimodeaction}
\end{align}
Each chiral boson $\ph_i$ has its own edge velocity $v_i$, a chirality $\eta_i$ and a coupling parameter $\kappa_i>0$. The index $i$ refers to the $i$'th chiral boson of the right-moving edge. The left moving edge consists of a similar set of bosons, but with opposite chiralities i.e. $\eta^L_i = -\eta^R_i$, etc. We will always work with the right moving current unless explicitly stated otherwise. It is possible to have $\kappa_i=0$, which corresponds to a chiral boson which does not couple to the electromagnetic field. Such a boson already falls into the category of neutral channels, so we assume $\kappa_i > 0$.

It is possible to formulate the edge theory in terms of coupled chiral bosons, which is usually done through use of a \mbox{$K$-matrix}\citep{wen1991,wen2004}. Starting from this formulation we can always switch to a different basis of fields through a linear transformation, which results in an action of the form Eq.~\eqref{multimodeaction}. Therefore there is no loss of generality by assuming decoupled chiral bosons.

For each boson we have the equation of motion
\begin{align}
(-\eta_i\partial_t - v_c \partial_x)\ph_i = -\kappa_i U_R~.
\end{align}
Since the channels are decoupled we can apply the same argument as before to obtain the edge current operator for each channel separately. The charge density, its corresponding conserved charge and the edge current density operator of the $i$'th channel are
\begin{align}
\rho_i &= \frac{\kappa_i}{2\pi} \partial_x\ph_i~,  \qquad
\mathcal{Q}_i = \frac{\kappa_i}{2\pi}\int_{\Sigma_R} \partial_x\ph_i~dx ~, \\
j_i &= -\frac{\kappa_i}{2\pi} \partial_t\ph_i  = \eta_i v_i \rho_i
- \eta_i\frac{\kappa_i^2}{2\pi} U_R~.
\end{align}
Likewise, the commutation relations also decouple
\begin{align}
[\partial_x\ph_i(x,t), \ph_j(0,0)] = i\eta_i 2\pi  \delta(x-\eta_i v_i t)\delta_{ij}~.
\end{align}

The total charge density, electric charge and edge current of the right moving edge is the sum of these operators
\begin{align}
\rho_R &= \sum_i \rho_i~, &
\mathcal{Q}_R &= \sum_i \mathcal{Q}_i~, &
j_R &= \sum_i j_i~.
\end{align}
A similar definition applies to the left moving edge. 

The total current operator is again the sum $j_R(x) + j_L(x)$, Eq.~\eqref{currentoperator}. To obtain the current-voltage relation \eqref{Hallrelation} we assume that each channel is in chemical equilibrium, meaning the density matrix is of the form $e^{-\beta K_0}/Z$ and the charge density of each channel vanishes $\langle \rho_i\rangle = 0$. The expectation value of the right-moving edge current is
\begin{align}
\langle j_R(x,t) \rangle = -\frac{1}{2\pi}U_R\sum_i \eta_i \kappa_i^2
\end{align}
and similarly for the left-moving edge current. For a right moving edge we require $\bigl(\sum_i \eta_i \kappa_i^2\bigr) > 0$, while for a left moving edge it is negative. The usual conductivity relation Eq.~\eqref{Hallrelation} is obtained provided we have
\begin{align}
\sum_i \eta_i \kappa_i^2 = \nu~.
\end{align}
This restriction is in fact a consequence of anomaly cancellation\citep{bieri2011}, so we assume that it holds. Unlike the single-channel case the conductivity does not uniquely specify the couplings $\kappa_i$ (recall that in single channel case we simply have \mbox{$\kappa_1 = \snu$}). To fully specify the topological order we also need to define the electron operators of the theory, which in turn determines the quasiparticle content. We refer to the literature for further discussions on this classification scheme.

A generic quasiparticle operator is of the form
\begin{align}
\psi^\dag_R(x,t) \propto \sigma_R(x,t)e^{-i\sum_i \eta_i q_i \ph_i(x,t)} \label{quasiparticlemulti}
\end{align}
which is defined by the $q_i$'s. The electric charge $Q$ of the quasiparticle is determined using the commutation relation with the charge operator
\begin{align}
Q\psi^\dag_R = [\mathcal{Q}_R, \psi^\dag_R] &= \frac{1}{2\pi}\sum_i \kappa_i \int [\partial_x\ph_i(x), \psi^\dag_R]~dx ~.
\end{align}
It follows that the charge is given by
\begin{align}
Q = \sum_i \kappa_i q_i~.
\end{align}
In addition the conformal dimension for the $i$'th channel is $h_i = \frac{q_i^2}{2}$ and so the total conformal dimension equals \mbox{$h = h_n + h_c$} with 
\begin{align}
h_c = \sum_i \frac{q_i^2}{2}~.
\end{align}
Finally, the non-equal time commutation relations between the current and the quasiparticle is given by
\begin{multline}
[j_R(x,t), \psi_R^\dag(y,t') ] = \\
\Bigl( \sum_i  \eta_i v_i \kappa_i q_i \delta(x-y - \eta_i v_i (t-t')) \Bigr) \psi_R^\dag(y,t')~.
\end{multline}
The generic form of the quasiparticle operator \eqref{quasiparticlemulti} involves all the channels of the edge theory, although this mixing does not always occur. 

An example of a state which is described by multiple charged chiral bosons is the Moore-Read trial state\citep{mooreread1991,greiter1991,milovan1996} of the $\nu = \frac{5}{2}$ plateau\citep{willett1987,pan1999}. Here we deal with a half-filled Landau level on top of two fully filled Landau levels. The edge theory consists of two chiral bosons with couplings \mbox{$\kappa_1 = \kappa_2 = 1$}, a third chiral boson with \mbox{$\kappa_3 = \frac{1}{\sqrt{2}}$} and a neutral channel described by the chiral Ising model. This corresponds to a \mbox{conductivity} of \mbox{$\nu = \frac{5}{2}$}. All channels are completely decoupled and have the same chirality. The quasiparticle operators do not mix different chiral bosons, so for each quasiparticle the sum appearing in Eq.~\eqref{quasiparticlemulti} consists of only one term.

A second example is a hierarchial trial state\citep{haldane1983b,halperin1984} of the \mbox{$\nu =\frac{2}{5}$} plateau. The trial state is formed through condensation of quasiparticles in the \mbox{$\nu = \frac{1}{3}$} state. The corresponding edge\citep{wen1992} consists of two (co-propagating) chiral bosons with couplings \mbox{$\kappa_1 = \frac{1}{\sqrt{3}}$} and \mbox{$\kappa_2 = \frac{1}{\sqrt{15}}$}, which brings the conductivity to~\mbox{$\nu=\frac{2}{5}$}. A simplified description assumes the distance between the two charged channels is large and the chiral bosons can be treated as completely decoupled. Each quasiparticle operator is then associated with strictly one chiral boson. 

In practice the distance between the channels is small, the Coulomb interaction needs to be taken into account\citep{wen1992} and the channels no longer decouple (although the currents still commute). In this case it is possible to diagonalize the interaction term through a linear transformation of the fields. The new fields are, again, completely decoupled. In this new basis the quasiparticle and electron operators are constructed from multiple fields, and in particular the sum appearing in \eqref{quasiparticlemulti} contains both chiral bosons of the new basis.

We finalize this discussion by noting that it is currently not completely clear if the case of counter propagating charge modes arises in the quantum Hall effect, as they have never been experimentally verified. One explanation for this is that counter propagating modes are unstable in the presence of disorder. In Ref.~\onlinecite{kane1994} it was found that for the $\nu=2/3$ state disorder induces tunnelling of charge between the counter propagating modes. This results in a different effective edge theory that consists of a single charged mode and a counter propagating \emph{neutral} mode. In this work we do not consider such dynamical effects which alter the edge theory away from the point contact. We simply assume the different channels completely decouple, and allow for the possibility of counter propagating modes. A recent experiment\citep{bid2010} suggests that counter propagating neutral modes are in fact present in multiple states, including the $\nu = 5/2$ state.

\section{Tunnelling point contact}\label{sec_tunnelling_pc}
\begin{figure}
\includegraphics[width=.45\textwidth]{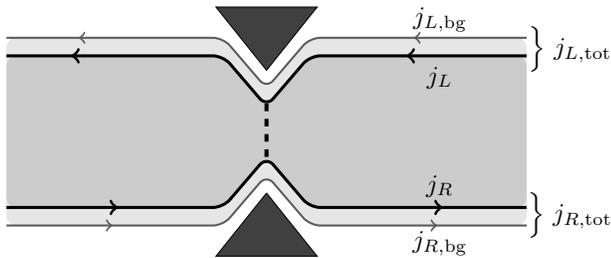}
\caption{The point contact induces tunnelling between the two edges. Tunnelling occurs between the inner channels of the edges. We decompose the total edge current $(j_{R/L,\text{tot}})$ into channels which are partially reflected ($j_{R/L}$), and which are fully transmitted ($j_{R/L,\text{bg}}$).} \label{fig:tunnellingsetup}
\end{figure}

\subsection{Tunnelling Hamiltonian and tunnelling current}\label{sec:tunham}
We consider a quantum Hall bar at filling fraction $\nu$ with two disconnected edges\citep{chamon1995,chamon1996,fendley2007}. A point contact acts as a restriction forcing opposite edges together thereby inducing tunnelling of quasiparticles between the edges as is sketched in Figure~\ref{fig:tunnellingsetup}. The tunnelling operator is the operator which tunnels a quasiparticle from the lower to the upper edge and it is defined as
\begin{align}
\V & = \psi_L^\dag(x=0)\psi_R(x=0)~. \label{eq:tunV}
\end{align}
Here $\psi$ is the quasiparticle operator defined in Eq.~\eqref{quasiparticlesingle} and Eq.~\eqref{quasiparticlemulti}. The quasiparticle is characterized by its quasiparticle charge $Q$ and conformal dimensions $h_n$ and $h_c$. In the multichannel case we assume the couplings $\kappa_i$ and individual charges $q_i$ are known. The tunnelling Hamiltonian is the tunnelling operator together with a tunnelling coupling constant
\begin{align}
\htu &= \Gamma \V + \Gamma^* \V^\dag~. \label{eqn:tunham}
\end{align}
It is treated as a perturbation to the grand canonical Hamiltonian $K_0$, Eq.~\eqref{eq:grandcanonicalHamiltonian}. 

In the presence of a voltage bias a net current of quasiparticles tunnels from one edge to the other, resulting in a tunnelling current. This is called the tunnelling or backscattering current $I_B$. It is defined as the rate of change of the charge difference between the edges. Using Heisenberg's equation of motion we obtain
\begin{align}
\hat{I}_B  \equiv \frac{e}{2}\frac{d}{dt}(\mathcal{Q}_L - \mathcal{Q}_R) &= -i\frac{e}{2}[\mathcal{Q}_L - \mathcal{Q}_R,K_0 + \htu]~.\label{eqn:backcurrent1}
\end{align}
Charge is conserved in the equilibrium system, so \mbox{$[K_0, \mathcal{Q}_{R/L}] = 0$}. The tunnelling operators are defined in terms of quasiparticle operators with charge $Q$, and so \mbox{$[\mathcal{Q}_L,\mathcal{V}] = -[\mathcal{Q}_R,\mathcal{V}] = Q\mathcal{V}$}. This also applies to the case of multiple charged channels. We have
\begin{align}
\hat{I}_B  = -iQe\left(\Gamma\V - \Gamma^*\V^\dag\right)~.\label{eqn:backcurrent2}
\end{align}

\subsection{Background current and multichannel case}\label{sec:background_current}
The point contact induces tunnelling of quasiparticles between the innermost channels of the left- and right moving edge. In particular it does not always involve all edge channels. An example is the Moore-Read state for the $\nu_{\text{tot}} = 5/2$ plateau. In this case the outer channels correspond to the fully filled Landau levels which are fully transmitted. Tunnelling occurs only between the inner channels described by the chiral Ising model times a chiral boson.

We therefore decompose the edge current into two pieces: the channels which are fully transmitted and not involved in the tunnelling process (called the background current), and the channels which are partially reflected (called the reflected current). This decomposition is sketched in Figure~\ref{fig:tunnellingsetup}. The corresponding current operators are denoted $j_{R,\text{bg}}$ for the background current and $j_R$ for the reflected current. The total edge current operator is written as
\begin{align}
j_{R,\text{tot}} &= j_R + j_{R,\text{bg}}\label{eq:totalcurrent}
\end{align}
In addition the reflected current and background current can also consist of multiple channels. Note also that the conductivity splits accordingly
\begin{align}
\nu_{\text{tot}} = \nu + \nu_{\text{bg}}~.
\end{align}
The decomposition \eqref{eq:totalcurrent} is reflected in the definition of the tunnelling Hamiltonian and the tunnelling current, Eq.~\eqref{eqn:tunham} and \eqref{eqn:backcurrent1}. The perturbation $\htu$ commutes with the current operators of the channels not involved in the tunnelling, i.e. \mbox{$[\mathcal{Q}_{R,\text{bg}}, \htu] = 0$} and so
\begin{align}
[\mathcal{Q}_{R,\text{tot}}, \htu] = 
[\mathcal{Q}_{R}, \htu]~.
\end{align}
We can therefore treat the background current as an equilibrium system unaffected by the perturbation.

\section{Non-equilibrium formalism }\label{sec:formalism}
\subsection{Formalism}
The presence of a point contact together with an applied voltage bias couples the upper and lower edges and forces the system out of equilibrium and we require a formalism that takes this into account. In a non-equilibrium formalism\cite{rammer2007} the tunnelling Hamiltonian is treated as a perturbation of the grand canonical Hamiltonian $K_0$. Initially at some time $t < t_0$ the perturbation is absent and the system is described by an equilibrium density matrix of the form
\begin{align}
w_0 \equiv w(t_0) &= e^{- K_0/k_BT}/Z~.\label{eqn:initialdensity}
\end{align}
We also denote $\langle \cdots\rangle$ as the expectation value with respect to  $w_0$, 
\begin{align}
\langle \cdot \rangle \equiv  \tr{w_0\cdots}~.
\end{align}
This density matrix further factorizes as a product of density matrices -- one for each channel of the system. At some time $t_0$ the perturbation is switched on adiabatically and the system is slowly driven away from equilibrium. Eventually, after the perturbation is fully switched on ($t\gg t_0$) the system is described by a steady state. In our approach we make use of the fact that (1) the initial state is an equilibrium state and (2) the unitary time evolution of the system is completely described by the (known) perturbed Hamiltonian $K = K_0 + \htu$.

Concretely, when the system reaches a steady state the expectation value of an operator $\mathcal{O}$ is given by \mbox{$\langle \mo(t) \rangle = \tr{w_0 \mathcal{O}_K(t)}$} where $\mathcal{O}_K(t) $ is the Heisenberg representation of the operator $\mathcal{O}$ with respect to the grand canonical Hamiltonian $K$,
\begin{align}
\mathcal{O}_K(t) = \mathcal{S}_{K}^\dag(t,t_0)\mathcal{O}_K(t_0)S_{K}(t,t_0)~.\label{eqn:expval}
\end{align}
The unitary time evolution operator $S_K(t,t_0)$ solves the Schr\"odinger equation,
\begin{align}
i\partial_t \mathcal{S}_{K}(t,t_0) = K \mathcal{S}_{K}(t,t_0)\label{app:schroedinger}
\end{align}
and $S_K(t,t) = 1$. Following Ref.~\onlinecite{rammer2007} we factorize the time evolution operator as \mbox{$\mathcal{S}_K(t,t_0) = e^{-iK_0(t-t_0)} \ut(t,t_0)$}. From Eq.~\eqref{app:schroedinger} it follows that the unitary operator $\ut(t,t_0)$ satisfies the equation of motion
\begin{align}
i\partial_t  \ut(t,t_0) &= \htu(t)  \ut(t,t_0)\\
\htu(t) &\equiv  e^{iK_0 t}\htu  e^{-iK_0t}~.
\end{align}
Here $\htu(t)$ is in an interaction-like picture with its time evolution dictated by the unperturbed Hamiltonian $K_0$. The time evolution operator $\ut$ is also known as the S-matrix operator and it is given by Dyson's series
\begin{align}
\ut(t,t_0)  &= \mathcal{T}\exp\bigl(-i\int_{t_0}^{t} \htu(t') ~dt'\bigr) \nonumber \\
&= 
 1 + \sum_{n=1}^\infty \frac{\left(- i\right)^n }{n!}  \Bigl[\prod_{i=1}^n\int_{t_0}^{t}dt_i \Bigr]~ \mathcal{T}\prod_{j=1}^n \htu(t_j) \label{app:timevoexp}
\end{align}
Here $\mathcal{T}$ is the time-ordering operator and the exponentiated form is an abbreviation for the corresponding expansion. Similarly, we set for an operator $\mathcal{O}$
\begin{align}
\mathcal{O}_{K_0}(t) = e^{iK_0 t} \mathcal{O} e^{-iK_0t}~.\label{eq:timedependence}
\end{align}
By using the factorization of the unitary time evolution operator in \eqref{eqn:expval} and taking the limit $t_0 \rightarrow -\infty$ we obtain for an operator $\mathcal{O}$ its expectation value
\begin{align}
\mathcal{O}^I(t)  &\equiv \ut^\dag(t,-\infty)\mathcal{O}_{K_0}(t)\ut(t,-\infty) \label{interactionpic} \\
\langle \mathcal{O}^I(t) \rangle &= \tr{w_0 O^I(t)} \nonumber\\
&= \tr{w_0\ut^\dag(t,-\infty)\mathcal{O}_{K_0}(t)\ut(t,-\infty)}~. \label{expval}
\end{align}
Here $\mathcal{O}^I(t)$ is still the Heisenberg representation $\mathcal{O}_K(t)$, but with the time evolution operators factorized. The superscript ${}^I$ denotes that the tunnelling Hamiltonian is switched on and the operator is taken in the Heisenberg representation. The effect of the perturbation $\htu$ is completely captured by the time evolution operator $\mathcal{U}$. All correlators are evaluated with respect to the equilibrium density matrix $w_0$. 

As an example the expectation value of the tunnelling current is given by
\begin{align}
I_B &= \langle \ib(t)\rangle\\ 
\ib(t) &=   \ut^\dag(t,-\infty)\hat{I}_B(t) \ut(t,-\infty)~.\label{eq:interactingbcurrent}
\end{align}
If we want to explicitly determine this correlator we need to resort to perturbation theory.

\subsection{A non-equilibrium Kubo formula}
In the formalism presented here the effect of the tunnelling perturbation is fully captured by the time evolution operator $\mathcal{U}(t,t_0)$. In linear response theory the time evolution operator Eq.~\eqref{app:timevoexp} is expanded to lowest order in the tunnelling coupling constant, which leads to the Kubo formula,
\begin{align}
\mathcal{O}^I(t) \cong \mathcal{O}_{K_0}(t)  - i\int_{-\infty}^t [\mathcal{O}_{K_0}(t), \htu(t')] ~dt' + \ldots~.
\end{align}
The dots represent higher order contributions. We present here an extension of the Kubo formula, which includes the higher order contributions. This non-equilibrium Kubo formula is given by\citep{fujii2010}
\begin{multline}
\mo^I(t)  = \mo_{K_0}(t)\\
  -i\int_{-\infty}^t \ut^\dag(t',-\infty)[\mo_{K_0}(t),\htu(t')]\ut(t',-\infty)~ dt' \label{eq:op}
\end{multline}
We emphasize that this expression is an operator identity. Ref.~\citep{fujii2010} obtains this formula for class of operators which commute with the equilibrium Hamiltonian $K_0$. The second term is the difference of the operator in a system in equilibrium and a system out of equilibrium,
\begin{align}
 \delta & \mo^I(t) \equiv \mo^I(t) - \mo_{K_0}(t)  \\
 &= -i\int_{-\infty}^t \ut^\dag(t',-\infty)[\mo_{K_0}(t),\htu(t')]\ut(t',-\infty)~ dt'\nonumber
\end{align}
This equation separates the effect of the perturbation on the operator $\mo$ when the perturbation is turned on and the system is forced out of equilibrium.

A proof of this relation is presented in appendix \ref{app1:kirchoff}. In this proof we apply the expansion of the time evolution operator Eq.~\eqref{app:timevoexp} to the operator in the interaction representation Eq.~\eqref{interactionpic}. Through some combinatorial manipulations of these expansions we recover the non-equilibrium Kubo formula \eqref{eq:op}.

\section{Edge current operator in the non-equilibrium formalism}\label{sec:edgecurrent}
In the absence of the point contact, the current through the system is given by the usual quantum Hall relation $\imax = \frac{\nu}{2\pi}(U_L-U_R)$. In the presence of a point contact this Hall relation no longer holds. The point contact induces a tunnelling current $I_B$, which is effectively a form of backscattering, since the edge currents of the system are chiral. On the basis of charge conservation we expect the current in the presence of a point contact to be
\begin{align}
\io = \imax - I_B~. \label{eqn:currentpert}
\end{align}
We now show that this relation is also satisfied at the level of the operators. For this we make use of the non-equilibrium Kubo formula. Recall that in the interaction representation the total current operator is
\begin{align}
\hat{I}_0^I(x,t) = j^I_R(x,t) + j_L^I(y,t)~. \label{eq:total_current_operator}
\end{align}
Here $j_R^I$ and $j_L^I$ are the edge currents in the interaction picture, Eq.~\eqref{interactionpic}. We focus initially on an edge with a single charged channel and comment on the multichannel case at the end of the section. 

We now apply the non-equilibrium Kubo formula Eq.~\eqref{eq:op}. For this we need the commutator of the edge current and the tunnelling Hamiltonian. We use the commutation relations of the edge current with the quasiparticle operator, Eq.~\eqref{eq:commutation_relation_qp_edge_current}, and the expression of the tunnelling Hamiltonian in terms of the quasiparticles, $\htu = \Gamma \psi_L^\dag \psi_R + \text{c.c.}$. This gives
\begin{align}
[j_{R}(x,t),\htu(t')] &= -i\eta_R v_c \hat{I}_B(t')\delta(x - \eta_R v_c(t-t')) \nonumber
\\
[j_{L}(x,t),\htu(t')] &= i\eta_L v_c \hat{I}_B(t')\delta(x -\eta_L v_c(t-t')) \label{eq:commutator_edge_hamilt}
\end{align}
with $\eta_R = +1$ and $\eta_L = -1$. Plugging this into \eqref{eq:op} for $j_{R/L}^I$ and performing the integration over $t'$ results in
\begin{align}
j_{R}^I(x,t) &= j_{R}(x,t) -\theta(x)\ib(t-x/v_c)~. \label{rightmoving} 
\\
j_{L}^I(x,t) &= j_{L}(x,t) -\theta(-x)\ib(t+ x/v_c)
\end{align}
Here $\theta(x)$ is the unit step function, and $j^I(x,t)$ and $\ib$ are the edge current and the tunnelling current operator in the interaction representation, see Eq.~\eqref{eq:interactingbcurrent}.

This operator has an intuitive meaning. It is a reflection of both charge conservation and the chiral structure of the edge current. Consider Eq.~\eqref{rightmoving} for the rightmoving current. For the region $x<0$ the operator reduces to $j_{R}^I(x,t) = j_{R}(x,t)$, meaning the current operator in this region is not affected by the presence of the tunnelling point contact. This is as expected, since the region $x<0$ is ``upstream" of the point contact. For the region $x>0$ the backscattering current $I_B$ at a retarded time $(t-x/v_c)$ is subtracted. The backscattering current is the charge transferred from the lower to the upper edge and is therefore subtracted from the current past the point contact (it is also subtracted from the left moving current because of the direction of total current). The identity resembles Kirchoff's law as charge is conserved along the point contact.

The fact that we subtract the operator $\ib$ from $j_R$ at a retarded time $t-x/v_c$ is a manifestation of the chiral and causal structure. Chirality and translational symmetry enforces all observables to be functions of the combination $t-x/v_c$. A similar argument is used in Ref.~\onlinecite{wang2011} as a derivation of the edge current operator for the system out of equilibrium. The chiral structure takes into account the position of the point contact (at $x_R=0$, hence the step function), the chirality of the edge (right-moving) and the finite velocity of the charged channel.

The total current operator in the interacting regime is now
\begin{align}
\hat{I}_0^I(x,t)  &= j_R(x,t)+j_L(x,t) -  \ib\bigl(t-|x|/v_c\bigr)~.   \label{eq:transcurrent}
\end{align}
This indeed reproduces the current relation Eq.~\eqref{eqn:currentpert}  
\begin{align}
\io &= 
\langle j_R(x,t)+j_L(x,t) \rangle - \langle \ib\bigl(t-|x|/v_c\bigr)  \rangle \nonumber\\
&= \imax - I_B~.
\end{align}
A similar relation applies to the charge density operators. When we apply the non-equilibrium Kubo formula to these operators we find
\begin{align}
\rho_R^I(x, t) & = \rho_R(x,t) - \frac{1}{v_c} \ib(t-x/v_c)\theta(x)\nonumber\\
\rho_L^I(x, t) & = \rho_L(x,t) + \frac{1}{v_c} \ib(t+x/v_c)\theta(-x) 
\end{align}
Note that the sign of $I_B$ in the equations are merely a consequence of our conventions (direction of the current and backscattering current, charge of the tunnelling quasiparticle, etc.)

Let us remark on the more general case of multiple charged channels. First note that the inclusion of background currents (see Section~\ref{sec:background_current}) does not modify the relation, since the background currents commute with the tunnelling Hamiltonian. This is intuitively clear, since the background currents are fully transmitted.

In the more general case the additional charged channels do not commute with the tunnelling Hamiltonian. The total edge current is a sum of the background currents plus the reflected edge currents 
\begin{align}
j_{R,\text{tot}} &= j_R + j_{R,\text{bg}}~, &
j_R &= \sum_{i} j_i~.
\end{align}
Each channel is characterised by its own edge velocity $v_i$ and chirality $\eta_i$. The commutator of the edge current operator with the tunnelling Hamiltonian becomes
\begin{multline}
[j_{R,\text{tot}}(x,t),\htu(t')] = [j_{R}(x,t),\htu(t')] =  \\
-i \hat{I}_B(t') \sum_i \frac{\kappa_i q_i}{Q} \eta_i v_i  \delta(x - \eta_i v_c(t-t')) ~.
\end{multline}
and for completeness we also note the left moving edge (with chiralities $\eta^L_i$)
\begin{multline}
[j_{L,\text{tot}}(x,t),\htu(t')] = [j_{L}(x,t),\htu(t')] =  \\
i \hat{I}_B(t') \sum_i \frac{\kappa_i q_i}{Q} \eta_i^L v_i  \delta(x + \eta^L_i v_c(t-t')) ~.
\end{multline}
The charge of the quasiparticle in this case is given by $Q = \sum_i \kappa_i q_i$. The edge current operator in the interaction picture is given by
\begin{multline}
j_{R,\text{tot}}^I(x,t) =\\ j_{R,\text{tot}}(x,t) -\sum_i \bigl(\frac{\kappa_i q_i}{Q}\bigr) \eta_i  \theta(\eta_i x)\ib(t-\eta_i x/v_i)~. \label{eq:current_noneq_multichannel}
\end{multline}
The summation reflects the chiral structure of each channel separately and the current relation Eq.~\eqref{eqn:currentpert} is again obtained.

\section{Non-equilibrium noise}\label{sec:noise}
\begin{figure}
\includegraphics[width=.35\textwidth]{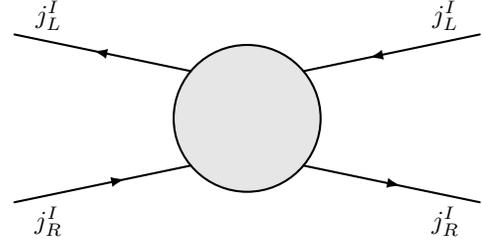}
\caption{The point contact viewed as a scattering source at $x_R = x_L = 0$ with the edges depicted as incoming and outgoing edge currents. The arrows denote the direction of the local electric current. The edge currents are taken in the interaction picture.}\label{fig:scatteringsource}
\end{figure}

The main result of the previous section is the operator identity Eq.~\eqref{eq:transcurrent} which captures the effect of the tunnelling Hamiltonian on the edge current. In this section we analyse the noise in the edge current in the non-equilibrium system. Using the identity Eq.~\eqref{eq:transcurrent} we can relate the noise in the edge current out of equilibrium to the noise in the tunnelling current. This results in a non-equilibrium fluctuation-dissipation theorem\citep{kanefisher1994} and an expression for the excess noise in the edge current. Put differently, we are studying the effects of the non-equilibrium Kubo formula on autocorrelators and their Fourier transform.

Let us first recall some definitions\citep{blanter2000, martin2005, clerk2010}. Given an operator $\mathcal{O}$ we set $\Delta \mathcal{O}(t) = \hat{\mathcal{O}}(t) - \langle \hat{\mathcal{O}}\rangle$ and define the autocorrelator $S_{\mathcal{O}}(t)$ as
\begin{align} 
S_{\mathcal{O}}(t) &= \langle \Delta \mathcal{O}(t)\Delta\mathcal{O}(0)\rangle \nonumber\\
 &=\langle \mathcal{O}(t)\mathcal{O}(0)\rangle - \langle \mathcal{O}\rangle^2~.
\end{align}
The non-symmetrized noise is the corresponding Fourier transform
\begin{align}
S_{\mathcal{O}}(\omega) = \int  e^{i\omega t} S_{\mathcal{O}}(t)~dt~. \label{noise_definition}
\end{align}
The symmetric and antisymmetric combinations of the noise are denoted by
\begin{align}
C_{\mathcal{O}}(\omega) &= \frac{1}{2}\left(S_{\mathcal{O}}(\omega) + S_{\mathcal{O}}(-\omega)\right) \\
R_{\mathcal{O}}(\omega) &= \frac{1}{2}\left(S_{\mathcal{O}}(\omega) - S_{\mathcal{O}}(-\omega)\right)~.
\end{align}
The same notation is used in Ref.~\onlinecite{kanefisher1994}.

\subsection{Noise in the outgoing edge current} \label{sec:noise_in_outgoing}
In the spirit of Ref.~\onlinecite{chamon1996} we think of the point contact as a scattering source with the edges as two incoming and two outgoing branches, see Figure~\ref{fig:scatteringsource}. We focus on the noise in the outgoing branch of the right-moving edge, which corresponds to the noise in $j_{R,\text{tot}}^I(x,t)$ for $x>0$. We first consider the case of a single reflected charged channel plus any number of background currents which are fully transmitted. The edge current operator is
\begin{multline}
\Delta j_{R,\text{out}}^I(x,t) = \\
 j^I_R(x,t)  + j_{R,\text{bg}}(x,t)- \langle j^I_R(x,t)+j_{R,\text{bg}}(x,t) \rangle  
\end{multline}
with $j^I_R$ given by \eqref{rightmoving} and $x>0$. The noise in this outgoing edge current is defined as
\begin{align}
S_{\text{out}}(t) &= \langle \Delta j_{R,\text{out}}^I(x,t) \Delta j_{R,\text{out}}^I(x,0)\rangle \nonumber\\
&= \langle \Delta j_R^I(x,t) \Delta j_R^I(x,0)\rangle \nonumber\\
&+ \langle \Delta j_{R,\text{bg}}(x,t) \Delta j_{R,\text{bg}}(x,0)\rangle, & &x>0~.
\end{align}
We now show that this non-equilibrium noise is completely determined in terms the noise of the tunnelling current and the equilibrium noise of the edge current. For this we substitute for $j_R^I$ the operator equation \eqref{rightmoving} and expand to obtain (momentarily suppressing the $x$-dependency of the edge current operator)
\begin{align}
S_{\text{out}}(t) &= \langle \Delta j_R(t) \Delta j_R(0)\rangle + \langle\Delta j_{R,\text{bg}}(t) \Delta j_{R,\text{bg}}(0)\rangle \\
&+ \langle \Delta I_B^I(t-x/v_c) \Delta I_B^I(-x/v_c) \rangle \nonumber\\
&- \langle \Delta j_R(t+x/v_c)  I_B^I(0) +  I_B^I(t-x/v_c)\Delta j_R(0)\rangle\nonumber
\end{align}
The term $\langle \Delta j_{R,\text{bg}}(t) \Delta I_B^I(-x/v_c)\rangle$ vanishes and the term $\langle \Delta j_{R}(t) \Delta I_B^I(-x/v_c)\rangle$ simplifies to $\langle \Delta j_{R}(t+x/v_c) I_B^I(0)\rangle$. This expression is an expansion of $(j_R + j_{R,\text{bg}} -I_B)^2$. Note that we assume all edge currents operators completely decouple, and so there are no cross-correlations between different channels appearing in this expansion. The Fourier transform of the autocorrelator is the noise of in the outgoing current. We have
\begin{align}
S_{\text{out}}(\omega) = S_{\text{bg}}(\omega) + S_{I_B}(\omega) - \Delta S(\omega)~. \label{eq:mainresult1}
\end{align}
These contributions correspond to the equilibrium noise $\langle j_R j_R\rangle + \langle j_{R,\text{bg}} j_{R,\text{bg}}\rangle$, the noise in the tunnelling current $\langle I_B I_B\rangle$ and the cross terms $\langle j_R I_B\rangle +  \langle I_B j_R\rangle$. 

To be more precise, the first term in Eq.~\eqref{eq:mainresult1} is given by
\begin{align}
S_{\text{bg}}(\omega)& = \int e^{i\omega t} \langle \Delta j_{R,\text{tot}}(x,t) \Delta j_{R,\text{tot}}(x,0)\rangle~ dt \nonumber\\
={}& \omega N(\omega) G~.
\end{align}
which represents the noise of the edge in the absence of a point contact. Here \mbox{$G = \frac{\nu_{\text{tot}}}{4\pi}$} is half the total conductivity and \mbox{$N(\omega) = \coth\bigl(\frac{w}{2k_BT}\bigr) +1$}. This was determined in Section~\ref{sec1:model}. This expression is known as the (non-symmetrized) Nyquist-Johnson noise.

The second term in Eq.~\eqref{eq:mainresult1} is
\begin{align}
S_{I_B}(\omega) = \int  e^{i\omega t} \langle \Delta \ib(t) \Delta  \ib(0) \rangle ~dt \label{eq:backnoise}
\end{align}
which is the noise of the tunnelling current. It is a type of non-equilibrium noise meaning it is not described in terms of the Nyquist-Johnson relation. 

The final term is a cross-term between the tunnelling and edge current
\begin{align}
\Delta S(\omega) ={}& \int e^{i\omega t} \Big(\phantom{+}\langle \Delta j_R(x,t+x/v_c)  I_B^I(0)\rangle \nonumber\\
&\phantom{\int e^{i\omega t} \Big(}{+} \langle I_B^I(0) \Delta j_R(x,-t+x/v_c)\rangle \Big)dt\nonumber\\
={}&N(\omega) R_{I_B}(\omega)~. \label{crossnoise}
\end{align}
This contribution arises due to the correlation between the (equilibrium) edge current and the tunnelling current. It is completely determined by the anti-symmetrized noise of the tunnelling current.

The final expression for $\Delta S(\omega)$ in Eq.~\eqref{crossnoise} requires some justification. We make use of a \emph{non-equilibrium Ward identity} to simplify the expression for the correlators $\langle \Delta j_R  I_B^I\rangle$ and $\langle  I_B^I  \Delta j_R\rangle$. This is explained in Appendix~\ref{app2:crosscorr}. Ward identities are identities imposed on correlations functions due to symmetries of the theory. In this case this is due to the $u(1)$ symmetry associated with conservation of electric charge. The Ward identity allows us to incorporate the effect of the inserted current operator~$j_R$ in the correlator $\langle \Delta j_R  I_B^I\rangle$, without explicitly determining these correlators. In particular, we do not need to specify the structure of the neutral mode since it decouples from the current operator.

The expression for the noise in the outgoing current Eq.~\eqref{eq:mainresult1} combined with the expression for the cross term Eq.~\eqref{crossnoise} is our first main result. It is, up to a an equilibrium contribution, completely determined by the noise in the tunnelling current $S_{I_B}$. This is not uprising, since fluctuations that arise in the tunnelling current $I_B$ end up in the edge current. However, $S_{\text{out}}(\omega) \neq S_{\text{bg}}(\omega) + S_{I_B}(\omega)$. A correction term $\Delta S$ arises due to the correlation between the edge current and the tunnelling current. 

An alternative way of writing the noise in the outgoing current, Eq.~\eqref{eq:mainresult1}, is by writing the noise in the tunnelling current in terms of its symmetric and antisymmetric components, \mbox{$S_{I_B}(\omega) = C_{I_B}(\omega) + R_{I_B}(\omega)$}. This gives
\begin{align}
S_{\text{out}}^{xc}(\omega) = C_{I_B}(\omega) - \coth(\tfrac{w}{2k_BT}) R_{I_B}(\omega)\label{eq:mainresult2}
\end{align}
where we have replaced the left-hand side by the \emph{excess noise} in the outgoing current
\begin{align}
S_{\text{out}}^{xc}(\omega) &\equiv S_{\text{out}}(\omega) - S_{\text{out}}(\omega,V=0) \nonumber\\
&= S_{\text{out}}(\omega) - S_{\text{bg}}(\omega)~.
\end{align}
By definition the excess noise is obtained by subtracting the \mbox{$V=0$} contribution from the noise. In Eq.~\eqref{eq:mainresult2} the right hand side vanishes at \mbox{$V=0$} due to the equilibrium fluctuation-dissipation theorem. We show this in the next section. Keep in mind that $S_{I_B}(\omega, V=0)$ does not vanish, but the combination appearing on the right hand side in Eq.~\eqref{eq:mainresult2} at $V=0$ does. The noise in the edge current at zero voltage is therefore simply the equilibrium noise \mbox{$S_{\text{out}}(\omega, V = 0) = S_{\text{bg}}(\omega)$}.

Finally, we note that the excess noise $S_{\text{out}}^{xc}(\omega)$ is symmetric as follows from the right hand side of Eq.~\eqref{eq:mainresult2}
\begin{align}
S_{\text{out}}^{xc}(\omega) = C_{\text{out}}^{xc}(\omega)~. \label{symmequalsnonsymm}
\end{align}
Since $R_{{\text out}}^{xc}(\omega) = 0$ we also obtain
\begin{align}
R_{{\text out}}(\omega) = R_{\text {bg}}(\omega)~. \label{antisymmnoise}
\end{align}

\subsection{Non-equilibrium fluctuation-dissipation theorem}\label{subsection_noneqfdt}
Equations \eqref{eq:mainresult1} and \eqref{eq:mainresult2} are non-equilibrium relations between the noise in the outgoing and tunnelling current. In equilibrium both sides of Eq.~\eqref{eq:mainresult2} are zero due to the equilibrium fluctuation-dissipation theorem (FDT). To analyse this further we recall the FDT for a system in equilibrium and some operator $\mathcal{O}$,
\begin{align}
C_{\mathcal{O}}^{{\text eq}}(\omega) - \coth(\tfrac{\omega}{2k_BT})R_{\mathcal{O}}^{{\text eq}}(\omega) = 0~.\label{eqn:fdt}
\end{align}
The fact that the left hand side of Eq.~\eqref{eq:mainresult2} does not vanish signals the non-equilibrium nature of the excess noise in the outgoing current.

The equilibrium FDT is a direct consequence of the Kubo-Martin-Schwinger condition\cite{rammer2007} satisfied by the autocorrelator $S^{{\text eq}}_{\mathcal{O}}(t)$. This condition states that a two-point correlator computed with respect to a thermal state satisfies
\begin{align}
\langle \hat{A}(t)\hat{B}(0)\rangle_{\text{eq}} = \langle \hat{B}(0) \hat{A}(t+i/k_BT) \rangle_{\text{eq}}~. \label{eq:kms}
\end{align}
For an autocorrelator evaluated at equilibrium $S^{{\text eq}}_{\mathcal{O}}(t)$ this gives
\begin{align}
S_{\mathcal{O}}^{{\text eq}}(-t) &= S^{{\text eq}}_{\mathcal{O}}(t - i/k_BT)\nonumber\\
S_{\mathcal{O}}^{{\text eq}}(-\omega) &= e^{-\omega/k_B T} S^{{\text eq}}_{\mathcal{O}}(\omega)~.\label{excessnoise}
\end{align}
This equation and Eq.~\eqref{eqn:fdt} are both known as the equilibrium fluctuation-dissipation theorem. 

The noise in the outgoing current $S_{\text{out}}(\omega)$ does not satisfy the equilibrium FDT and is therefore a type of non-equilibrium noise. However, some terms appearing in its expansion Eq.~\eqref{eq:mainresult1} do. In particular the noise in the background current $S_{\text{bg}}$ and the correction term $\Delta S$ both satisfy the FDT. For $\Delta S(\omega)$ this follows from simply inserting Eq.~\eqref{crossnoise} into Eq.~\eqref{eqn:fdt}.

With these results we apply the equilibrium FDT to the first main result \eqref{eq:mainresult1} (the expansion of the noise in the outgoing edge current) and arrive at a \textit{non-equilibrium fluctuation-dissipation theorem} (NE-FDT), satisfied by the noise in the tunnelling and outgoing currents,
\begin{multline}
C_{{\text out}}(\omega) - \coth(\tfrac{\omega}{2k_BT})R_{{\text out}}(\omega) = \\C_{I_B}(\omega) - \coth(\tfrac{\omega}{2k_BT})R_{I_B}(\omega)~.\label{noneqfdt}
\end{multline}
This relation was derived by Kane and Fisher \citep{kanefisher1994} for a system of a chiral Luttinger liquid. In the case of Ref~\onlinecite{kanefisher1994} the noise in the tunnelling current is identified with the noise in the voltage drop over the point contact through $\hat{V}_B = \nu\frac{e^2}{h}\hat{I}_B$. As Kane and Fisher put it, this equation shows that the fluctuations in the edge and tunnelling currents are locked together. Here we have shown how this naturally follows from analysing the edge current operator in the non-equilibrium system. 

This NE-FDT relation is our second main result. Here we have generalized the proof to general fractional quantum Hall states, including non-Abelian states. The result also applies to the multichannel case, as we show in Section~\ref{sec:multichannel_noise}. The relation is a direct consequence of conservation of charge and the chirality of the edges. We emphasize though that the main result of this work is the expansion for the noise in the outgoing current Eq.~\eqref{eq:mainresult1} and the excess noise Eq.~\eqref{eq:mainresult2}, and these results do not follow from the NE-FDT.

\subsection{Shot noise limit}
A shot-noise relation is obtained in the zero frequency limit, $\omega\downarrow 0$. This relation for the excess noise is given by\cite{kanefisher1994, fendley1995b, wang2011, safi2008}
\begin{align}
S_{\text{out}}^{\text{xc}}(0) = S_{I_B}(0) - 2k_BT \frac{dI_B}{dV}~.\label{eqn:shotnoise}
\end{align}
To obtain this we use the relation
\begin{align}
\lim_{ \omega \downarrow 0} \coth(\tfrac{\omega}{2 k_BT})R_{I_B}(\omega)& =  2k_BT\frac{dI_B}{dV}~.\label{eq:diffcond}
\end{align}
Here $\frac{dI_B}{dV} = \frac{d}{dV} \langle \ib \rangle$ is the differential conductance of the tunnelling current. To prove \eqref{eq:diffcond} requires more work. First note that 
\begin{align}
\lim_{ \omega \downarrow 0} \coth(\tfrac{\omega}{2 k_BT})R_{I_B}(\omega)& = 2k_BT\frac{d S_{I_B}(\omega)}{d\omega}\Big|_{\omega = 0}
\end{align}
Next we show how you can prove that $\frac{d}{dV} \langle \ib \rangle$ equals $\frac{d S_{I_B}(\omega)}{d\omega}\Big|_{\omega = 0}$. For this we use the expression for $\ib$ in terms of the time evolution operator $\mathcal{U}$, Eq.~\eqref{eq:interactingbcurrent}, and the expansion of $\mathcal{U}$, Eq.~\eqref{app:timevoexp}. Acting with $\frac{d}{dV}$ on $\mathcal{U}$ results in
\begin{align*}
\frac{d}{dV} \ut(0,-\infty) &= 
-i\int_{-\infty}^0\mathcal{T}\Bigl(
\frac{d}{dV}\htu(t)e^{-i\int_{-\infty}^0\htu(t')dt'}\Bigr)dt
\\
&=-\frac{d}{d\omega}\int_{-\infty}^0 e^{i\omega t} \ut(0,-\infty)  \ib(t)  dt \Big|_{\omega=0}~. 
\end{align*}
Here we made use of
\begin{align}
\frac{d}{dV} \htu(t) &= 
-\frac{it}{2}[\mathcal{Q}_L - \mathcal{Q}_R, \htu(t)] \nonumber\\
&=-i \frac{d}{d\omega} e^{i\omega t}\hat{I}_B(t)\Big|_{\omega=0}~. 
\end{align}
By applying this relation to $\frac{dI_B}{dV}  = \frac{d}{dV}\langle \mathcal{U}^\dag\hat{I}_B\mathcal{U} \rangle$ we can relate the differential conductance to the noise
\begin{align}
\frac{dI_B}{dV}
&=\frac{d}{d\omega}\int_{-\infty}^0 e^{i\omega t} \langle \ib(t) \ib(0) - \ib(0) \ib(t)\rangle \Big|_{\omega = 0}\nonumber\\
&= \frac{d S_{I_B}(\omega)}{d\omega}\Big|_{\omega = 0}~. \label{differential_conductance}
\end{align}
Putting everything together results in the shot noise relation Eq.~\eqref{eqn:shotnoise}.

\subsection{The multichannel case} \label{sec:multichannel_noise}
In the interaction representation the edge current operator in the multichannel case is given by (Eq.~\eqref{eq:current_noneq_multichannel})
\begin{multline}
j^I_{R,\text{total}}  = j_{R,\text{bg}} + \sum_i j_i \\
-\sum_i \bigl(\frac{\kappa_i q_i}{Q}\bigr) \eta_i  \theta(\eta_i x)\ib(t-\eta_i x/v_i) \label{eq:current_noneq_multichannel2}
\end{multline}
The autocorrelator  of the total edge current is this operator squared. Since all channels decouple the autocorrelator is also a sum over the individual channels. Using the current relation Eq.~\eqref{eq:current_noneq_multichannel2} we expand this autocorrelator to 
\begin{multline}
S_{\text{out}}(t) = 
S_{\text{bg}}(t)+ S_{I_B}(t) \Bigl(\sum_{i}\theta(\eta_i x) \frac{\kappa_i q_i}{Q}\Bigr)^2    \\
 - \sum_{i,j} \theta(\eta_j x)\frac{ \kappa_j q_j}{Q}  \Delta S_{ij}(t)
\end{multline}
where $S_{\text{bg}}(t)$ is the autocorrelator of the total edge current in equilibrium $j_{R,\text{tot}}$, $S_{I_B}(t)$ is the autocorrelator of the tunnelling current, and
\begin{multline}
\Delta S_{ij}(t) = \langle\Delta j_{i}(x,t+\eta_j x/ v_j) \ib(0)  \rangle \\  +
\langle  \ib(0) \Delta j_{i}(x,-(t+\eta_j x/ v_j)) \rangle ~.
\end{multline}
The expression $\Delta S_{ij}(t)$ can be simplified using a non-equilibrium Ward identity which holds for each edge channel separately, see Appendix~\ref{app2:crosscorr}. For the diagonal components ($\Delta S_{ii}$) we obtain the same result as in the single-channel case, Eq.~\eqref{crossnoise}. For the off-diagonal components ($\Delta S_{ij}$ with $i\neq j$) some care is required since the velocities are assumed to be different. We find
\begin{multline}
S_{\text{out}}(\omega) = S_{\text{bg}} (\omega)  + S_{I_B}(\omega)\Bigl(\sum_{i}\theta(\eta_i x) \frac{\kappa_i q_i}{Q}\Bigr)^2  \\  
 - \Delta S(\omega) \Bigl(\sum_{i , j} \theta(\eta_i x)\frac{\kappa_i q_i}{Q} \frac{\kappa_j q_j}{Q} e^{i\omega x(\frac{\eta_i}{v_i} - \frac{\eta_j}{v_j})}\Bigr)\label{eq:excess_noise_multichannel}
\end{multline}
The functions $S_{\text{bg}}$, $S_{I_B}$ and $\Delta S(\omega)$ are the same as for the single channel case, see Section~\ref{sec:noise_in_outgoing}. The tunnelling current mixes different channels, which manifests itself in expression \eqref{eq:excess_noise_multichannel} through the oscillating contributions. This mixing enters the expression through an oscillating contribution which oscillates at a frequency \mbox{$x(\frac{\eta_i}{v_i} - \frac{\eta_j}{v_j})$} for each pair of channels as a function of varying $\omega$. For frequencies smaller compared to $v_i/x$ these phase factors are unity. The noise relation Eq.~\eqref{eq:excess_noise_multichannel} automatically takes into account the chirality of the edges and the effect of counter propagating modes. 

The nonequilibrium FDT that follows from Eq.~\eqref{eq:excess_noise_multichannel} is given by
\begin{multline}
C_{{\text out}}(\omega) - \coth(\tfrac{\omega}{2k_BT})R_{{\text out}}(\omega) = \\
\left(C_{I_B}(\omega) - \coth(\tfrac{\omega}{2k_BT})R_{I_B}(\omega)\right)
\Bigl(\sum_{i}\theta(\eta_i x) \frac{ \kappa_i q_i}{Q}\Bigr)^2 ~.\label{noneqfdt_multichannel}
\end{multline}
When all edge currents are co-propagating we have \mbox{$\sum_{i}\frac{ \kappa_i q_i}{Q}=1$}. The extra factor in Eq.~\eqref{noneqfdt_multichannel} compared to Eq.~\eqref{noneqfdt} only arises when we deal with counter propagating charged channels. The reason for this discrepancy is that the distinction of incoming and outgoing edge currents is not applicable for a system with counter propagating charged edge modes. If the left moving edge is taken into account we recover the usual NE-FDT.

The shot noise limit is given by
\begin{multline}
S_{\text{out}}(0) = S_{\text{bg}}(0) + S_{I_B}(0)\Bigl(\sum_{i}\theta(\eta_i x) \frac{ \kappa_i q_i}{Q}\Bigr)^2
 \\ +2k_BT \frac{dI_B}{dV} \Bigl(\sum_{i , j} \theta(\eta_i x)\frac{ \kappa_i q_i}{Q} \frac{ \kappa_j q_j}{Q}\Bigr)~.
\end{multline}

\section{Cross- and autocorrelators of edge currents}\label{app:othernoise}
\subsection{Edge current correlations}
In this section we expand on our previous results and investigate the finite frequency noise between the different branches of a quantum point contact. Following Ref.~\onlinecite{chamon1996} the starting point is the definition of the different branches of a quantum Hall point contact, as given by Figure~\ref{fig:scatteringsource}. We label these as $j_k(t) \equiv j_{R/L}(x_k, t)$ with $k=1,2,3,4$. These correspond to the different in- and outgoing edge currents. When we apply the non-equilibrium Kubo formula we obtain
\begin{align}
j_1^I(t) &= j_{R,\text{tot}}(x_1,t)  & &x_1 <0 \nonumber\\
j_2^I(t) &= j_{L,\text{tot}}(x_2,t)  & & x_2 > 0\nonumber\\
j_3^I(t) &= j_{R,\text{tot}}(x_3,t)- \ib(t-x_3/v_c)& &x_3 >0  \nonumber\\
j_4^I(t)& = j_{L,\text{tot}}(x_4,t)- \ib(t+x_4/v_c) & &x_4 <0  \label{app:branches} 
\end{align} 
We define the correlation between the $n$'th and $m$'th branch as
\begin{align}
\mathcal{S}_{nm}(\omega) = \int  e^{i\omega t} \langle \Delta j_n^I(t) \Delta j_m^I(0)\rangle~dt~.
\end{align}
It is now a straightforward process of determining all relations by inserting the current operators and simplifying all the terms. All autocorrelators decompose into terms already encountered in the main part of this paper and Appendix~\ref{app2:crosscorr}. Here we list them once more (we use $\eta = \pm$ to denote the right $(\eta = -)$ and left moving $(\eta = +)$ current),
\begin{align}
S_{\text{bg}}(\omega) &= \int e^{i\omega t} \langle \Delta j_{\eta,\text{tot}}(x,t) \Delta  j_{\eta,\text{tot}}(x,0) \rangle ~ dt  \nonumber\\
&= \omega N(\omega) G_{\text{tot}} \\
S_{I_B}(\omega) &=  \int  e^{i\omega t} \langle\ib(t)\ib(0)\rangle ~ dt \\
\mathcal{F}(\omega)& = \int  e^{i\omega t}  \langle \Delta j_\eta(x,t) \ib(\eta x/v_c) \rangle ~ dt \nonumber\\
&=  \frac{1}{2} N(\omega) \bigl(R_{I_B}(\omega)+iQ^2\langle \htu^I\rangle\bigr)\\
\Delta S(\omega) &= \mathcal{F}(\omega) + e^{\omega/T}\mathcal{F}(-\omega)\nonumber\\
&= N(\omega)R_{I_B}(\omega)~.
\end{align}
with $R_{\mathcal{O}}(\omega)$ the antisymmetric part of $S_{I_B}(\omega)$. Note also the relations
\begin{align}
\mathcal{F}(\omega)^* &=e^{\omega/T}\mathcal{F}(-\omega) \nonumber\\
2\text{Re}\left[\mathcal{F}(\omega)\right] &= \Delta S(\omega) \nonumber\\
2\text{Im}\left[\mathcal{F}(\omega)\right]& = Q^2 N(\omega) \langle \htu^I\rangle~.
\end{align}
The correlator $\langle \htu^I(0)\rangle$ arises as a consequence of the non-equilibrium Ward identity. Furthermore, we also have
\begin{multline}
S_{I_B}(\omega) - \Delta S(\omega) = \\
C_{I_B}(\omega) - \coth\bigl(\frac{\omega}{2 k_BT}\bigr) R_{I_B}(\omega)~.
\end{multline}

The diagonal terms of the correlation matrix $\mathcal{S}$ are
\begin{align}
\mathcal{S}_{11}(\omega)& = \mathcal{S}_{22}(\omega) = S_{\text{bg}}(\omega) \\
\mathcal{S}_{33}(\omega)& = \mathcal{S}_{44}(\omega) = S_{\text{bg}}(\omega) + S_{I_B}(\omega) - \Delta S(\omega) 
\end{align}
These autocorrelators are the noise of the edge currents. $\mathcal{S}_{33}$ and $\mathcal{S}_{44}$ are treated extensively in Section~\ref{sec:noise} and correspond to the noise in the outgoing branches. The correlations in the incoming branches, $\mathcal{S}_{33}$ and $\mathcal{S}_{44}$, are equilibrium noise due to the chirality of edge. 

The remaining correlators $\mathcal{S}_{nm}$ ($n\neq m$) cannot be interpreted as noise. Since $\mathcal{S}_{nm} = \mathcal{S}_{mn}^*$ we only look at the cases where $m>n$. We obtain
\begin{align}
\mathcal{S}_{12}(\omega)& =  0 \nonumber\\
\mathcal{S}_{34}(\omega)& =  e^{i\omega(x_3+x_4)/v_c} \left(S_{I_B}(\omega) - \Delta S(\omega)\right)\nonumber\\
\mathcal{S}_{13}(\omega)& = e^{i\omega(x_1-x_3)/v_c} \left(S_{\text{bg}}(\omega)  -\mathcal{F}(\omega)\right) \nonumber\\
\mathcal{S}_{24}(\omega)& = e^{i\omega(x_2-x_4)/v_c} \left(S_{\text{bg}}(\omega)  -\mathcal{F}(\omega)\right) \nonumber\\
\mathcal{S}_{14}(\omega)& = -e^{i\omega(x_1+x_4)/v_c} \mathcal{F}(\omega) \nonumber\\
\mathcal{S}_{23}(\omega)& = -e^{-i\omega(x_2+x_3)/v_c} \mathcal{F}(\omega)  ~.\label{correlations_general}
\end{align}
Naturally the incoming edge currents are not correlated, hence $\mathcal{S}_{12} = \mathcal{S}_{21}=  0$. The remaining correlators all contain phase factors which depend on the relative distance of the points of measurements to the point contact. 

\subsection{Edge currents noise and FDT's}

\begin{figure}
\includegraphics[width=.45\textwidth]{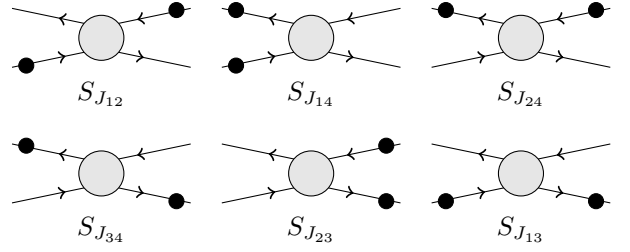}
\caption{The noise $S_{J_{nm}}$ represented pictorially. The operator $J_{nm}$ corresponds to the sum $j_n +j_m$ of edge current operators, and $S_{J_{nm}}$ is the corresponding noise. The figures represents the possible combinations of $j_n$ and $j_m$ (with $n\neq m$).}\label{fig:scatteringsource_multi}
\end{figure}

The cross correlations $\mathcal{S}_{nm}$ do not correspond to a type of noise when $n\neq m$. They do enter the expressions for the noise in operators which compare different edge currents. Such an operator is defined as
\begin{align}
J_{nm}(t) = j_n(t) + j_m(t),\quad n\neq m~.
\end{align}
The corresponding noise is given by
\begin{align}
S_{J_{nm}}= \mathcal{S}_{nn} + \mathcal{S}_{mm} + \mathcal{S}_{nm}+ \mathcal{S}_{mn}
\end{align}
We note the symmetry $J_{nm} = J_{mn}$ and set $m>n$. This gives rise to six different autocorrelators, which are depicted in Figure~\ref{fig:scatteringsource_multi}. We also assume the frequency $\omega$ at which the noise is measured is small compared to the combinations $v_c/x_{ij} = v_c/(x_i \pm x_j)$ as they appear in Eq.~\eqref{correlations_general}, and the noise is measured relatively close to the point contacts. In this limit there are four different cases for the cross-correlator noise. We first have $S_{J_{12}}$ and $S_{J_{34}}$,
\begin{align}
S_{J_{12}}(\omega) &= 2S_{\text{bg}}(\omega)  \\
S_{J_{34}}(\omega) &= 2S_{\text{bg}}(\omega) + 4S_{I_B}(\omega) - 4\Delta S(\omega)
\nonumber~.
\end{align}
Next we have $S_{J_{14}} = S_{J_{24}}$, where
\begin{align}
S_{J_{14}}(\omega) &= 2S_{\text{bg}}(\omega) + S_{I_B}(\omega) -2 \Delta S(\omega)
\end{align}
And finally $S_{J_{24}} = S_{J_{13}}$ with
\begin{align}
S_{J_{13}}(\omega) &=4S_{\text{bg}}(\omega)+ S_{I_B}(\omega) - 2\Delta S(\omega)  
\end{align}
All cross correlations are expressed in terms of the equilibrium noise of the background current $S_{\text{bg}}$ and the noise in the tunnelling current $S_{I_B}$ (since $\Delta S$ is also determined by $S_{I_B}$). In addition all these autocorrelators satisfy the same non-equilibrium FDT
\begin{multline}
C_{J_{nm}}(\omega) - \coth(\tfrac{\omega}{2 k_BT})R_{J_{nm}}(\omega) = \\
C_{I_B}(\omega) - \coth(\tfrac{\omega}{2 k_BT})R_{I_B}(\omega)~.
\end{multline}
In the shot noise limit we replace $S_{bg} \rightarrow 2k_BT G$ (with $G = G_{\text{tot}}$) and
\begin{align}
\lim_{\omega\downarrow 0} \Delta S(\omega) &=   2k_BT\frac{dI_B}{dV}\nonumber
\end{align}
This results in
\begin{align}
S_{J_{12}}(0) &= 4k_BT G \\
S_{J_{34}}(0) &= 4k_BT G  + 4S_{I_B}(0) - 8k_BT\tfrac{dI_B}{dV} \nonumber\\
S_{J_{14}}(0) = S_{J_{23}}(0) &=4k_BT G + S_{I_B}(0) - 4k_BT\tfrac{dI_B}{dV}\nonumber\\
S_{J_{13}}(0) = S_{J_{24}}(0) &=8k_BT G  + S_{I_B}(0) - 4k_BT\tfrac{dI_B}{dV}\nonumber
\end{align}

\section{Relation to experiments}
In general, the expressions we have obtained relate the noise or cross correlations in the outgoing edge currents to the noise in the tunnelling current. What all these relations have in common is that the noise in the edge currents is fully determined in terms of the tunnelling current ($I_B$), the noise in the tunnelling current ($S_{I_B}$) and equilibrium noise. In particular, $S_{I_B}$ and $I_B$ fully captures the non-equilibrium contribution to the noise in the edge current. We note that in this work we do not actually calculate $S_{I_B}$ or $I_B$. We refer to the literature for both perturbative and non-perturbative approaches towards this problem \citep{buttiker1992, kanefisher1994, chamon1995, chamon1996, fendley1995c, fendley1996, chang2003, trauzettel2004, martin2005,bena2006, bena2007,safi2008, ferraro2008,ferraro2010, ferraro2010a, ferraro2010b, carrega2011, iyoda2011, carrega2012, wang2011, wang2013}.

Still, the relations presented here have an important experimental interpretation. Experiments that measure shot noise \emph{do not} measure the shot noise of the tunnelling current directly. This can be seen from the experimental setups that are used in Refs.~\onlinecite{saminadayar1997,depicciotto1997,glattli1998,reznikov1999, glattli2000, griffiths2000, heiblum2000, comforti2002, chung2003a, chung2003b, heiblum2003, heiblum2006, dolev2008,chen2009,bid2009, dolev2010,dolev2010b, dolev2011, dolev2011b}. Instead, these experiments measure the shot noise of the edge current. Here we emphasize once more that the shot noise of the tunnelling current does not ``add up'' to the shot noise of the edge current. Eq.~\eqref{eqn:shotnoise} relates the two types of noise.

\section{Conclusion} \label{sec:conclusion}
In this work we investigated the relation between the noise in the outgoing edge current and the noise in the tunnelling current. We found an expression for the finite frequency (non-symmetrized) noise of the outgoing edge current, in terms of the noise in the tunnelling current and the equilibrium Nyquist-Johnson noise. From this we also find an expression for the corresponding excess noise in the edge current. This excess noise is symmetric in frequency and completely determined by the noise in the tunnelling  current. Finally, we also find a relation for the shot noise in the edge current.

Our approach made use of two new tools, which are also derived in this work. The first is the non-equilibrium Kubo formula. This operator equation separates the effect of time evolution due to a perturbation from the time evolution due to the free Hamiltonian. More specifically, in our context we obtain an equation relating the edge current operator for the system out of equilibrium, to the edge current operator for the system in equilibrium minus the tunnelling current, Eq.~\eqref{rightmoving}. This is an operator-version of Kirchoff's law and reflects charge conservation and the chiral structure of the edge theory. 

The second tool we made use of (and proved in the appendix) is a non-equilibrium Ward identity. This identity extends the equilibrium Ward identity to hold for certain correlators evaluated out-of-equilibrium. 

Our proof applies to generic quantum Hall edges consisting of a single chiral channel and any number of neutral channels. We have also extended the relation to apply to edges with multiple charged channels, possibly counter-propagating.

\section*{Acknowledgements}
J.K.S. and O.S. were supported by Science Foundation Ireland Principal Investigator award 08/IN.1/I1961 and 12/IA/1697. S.H.S. acknowledges support from EPSRC grants EP/I031014/1 and EP/I032487/1. 

\appendix

\section{NON-EQUILIBRIUM KUBO FORMULA}\label{app1:kirchoff}
In this appendix we will prove the non-equilibrium Kubo formula Eq.~\eqref{eq:op}, which is the operator equation \eqref{app:gen-result}. We define the Hamiltonian as $K(t) = K_0 +  \lambda(t)\htu$ with $\htu$ some perturbation that is adiabatically switched on at $t_0\rightarrow -\infty$ through the function $\lambda(t)$. We recall Dyson's series expansion of the S matrix operator $\ut$ given by
\begin{align}
\ut(t,t_0) &= \mathcal{T}\exp\left(-i\int_{t_0}^t  \lambda(t')\htu(t')~dt'\right)\nonumber\\
&= 1 + \sum_{n=1}^\infty \frac{\left(- i\right)^n }{n!} \int_{t_0}^{t}\dint{n}~ \mathcal{T}\prod_{j=1}^n \lambda(t_j)\htu(t_j)
\end{align}
Here $\htu(t) \equiv e^{iH_0 t} \htu(t_0)e^{-iH_0 t}$ and we use the notation $\int_{t_0}^{t}\dint{n} \equiv \prod_{i=1}^n \int_{t_0}^{t}dt_i$. For an operator $\mathcal{O}$ we set
\begin{align}
\mo(t) =   e^{iK_0 t} \mo(0)e^{-iK_0 t}\nonumber
\end{align}
where $\mo(0)$ is the Schroedinger representation of the operator. The expectation value of $\mathcal{O}$ is given by
\begin{align}
\langle \ut^\dag(t,t_0) \mathcal{O}(t)\ut(t,t_0) \rangle & = \tr{w_0\ut^\dag(t,t_0) \mathcal{O}(t)\ut(t,t_0)} \nonumber\\
w_0 & \equiv \frac{1}{Z}e^{-\beta K(t_0)} = \frac{1}{Z}e^{-\beta K_0} \nonumber
\end{align}
and the initial condition is a thermal state with respect to the Hamiltonian $K_0$. The operator equation we now prove relates the operator in the non-equilibrium regime to the equilibrium regime,
\begin{multline}
\ut^\dag(t,t_0)\mo(t) \ut(t,t_0) = \mo(t)\\
 -i \int_{t_0}^t ~ \lambda(t')\ut^\dag(t',t_0)[\mo(t), \htu(t')] \ut(t',t_0)~dt'\label{app:gen-result}
\end{multline}
In the main text we take $\lambda(t)\rightarrow 1$ and $t_0\rightarrow -\infty$. We assume these limits can be taken and the resulting identity holds.

The second term in Eq.~\eqref{app:gen-result} is the contribution due to the perturbation $\htu$, 
\begin{multline} 
\delta \mo(t) \equiv  \ut^\dag(t,t_0)\mo(t) \ut(t,t_0) - \mo(t)\\
=  -i\int_{t_0}^t  \lambda(t')\ut^\dag(t',t_0)[\mo(t),\htu(t')] \ut(t',t_0)~dt'~.
\end{multline}

To prove the result \eqref{app:gen-result} we start with the following expansion which follows automatically from the Dyson's series of $\ut$, 
\begin{align}
\ut^\dag(t,t_0)\mo(t) \ut(t,t_0) = \mo(t) + \sum_{n=1}^\infty \frac{(-i)^n}{n!} A^{(n)},\label{eqn:expansion}
\end{align}
where
\begin{align}
 A^{(n)} &=  \sum_{m=0}^n (-1)^{m}\binom{n}{m}  \label{eqn:summation} \\
 \int_{-t_0}^{t}&\dint{n} \bar{\mathcal{T}} \Bigl[\prod_{j=1}^m \htu(t_j)\Bigr] \mo(t)  \mathcal{T}\Bigl[\prod_{j=m+1}^n \htu(t_j)\Bigr]~.\nonumber
\end{align}
Here $\mathcal{T}$ and $\bar{\mathcal{T}}$ are time and anti-time ordering symbols, respectively, and they only act on the operators within the brackets. Furthermore, recall that $\int_{t_0}^{t}\dint{n} = \left[\prod_{i=1}^n\int_{t_0}^{t}dt_i\right]$ and we set empty products equal to one, i.e. $\prod_{j=n+1}^{n}\htu(t_j) \equiv 1$. For simplicity we have absorbed $\lambda(t)$ into the definition of $\htu$. 

Each summation $A^{(n)}$ can be written as a sum over commutators $\left[\mathcal{O}(t),i\htu(t_i)\right]$. First notice that if we exclude the effect of the remaining (anti-)time ordering but include the multiplicity due to the binomial $\binom{n}{m}$, the sum $A^{(n)}$ contains $2^n$ terms. This sum can be written as a sum over $2^{n-1}$ commutators. To illustrate this we fix the time ordering. The $m = 0$ and $m=1$ terms combine as
\begin{small}
\begin{multline}
 \Bigl(\begin{matrix}n \\ 0\end{matrix}\Bigr)\htu(t_1)\prod_{j=2}^n  \htu(t_j) 
-  \Bigl(\begin{matrix}n \\ 1\end{matrix}\Bigr) \htu(t_1)\mo(t)\prod_{j=2}^n \htu(t_j)=\\
 \Bigl(\begin{matrix}n-1 \\ 0\end{matrix}\Bigr)\left[\mo(t),\htu(t_1)\right]\prod_{j=2}^n \htu(t_j) \\
 -  \Bigl(\begin{matrix}n-1 \\ 1\end{matrix}\Bigr) \htu(t_1)\mo(t)\prod_{j=2}^n \htu(t_j)~.
\end{multline}
\end{small}
The first term contains the desired commutator. The second term can be combined with the $m=2$ contribution in \eqref{eqn:summation}. The remainder of this can be combined with the $m=3$ term, etc. The process is iterated until all terms are combined into commutators. The multiplicity of the $k$'th term in this sum over commutators is
\begin{align}
\sum_{m=0}^k (-1)^{m} \binom{n}{m} = (-1)^{k}\binom{n-1}{k}~.
\end{align}
To write down an expression of $A^{(n)}$ we need to incorporate the effect of time ordering. For that we fix the dummy indices such that $\left\lbrace t_1,\ldots,t_{m-1}\right\rbrace > t_m > \left\lbrace t_{m+1},\ldots,t_n\right\rbrace$, and relabel $t_m\rightarrow t'$ and $\left\lbrace t_{m+1},\ldots,t_n\right\rbrace\rightarrow \left\lbrace t_{m},\ldots,t_{n-1}\right\rbrace$. This can always be accomplished through relabelling of the integration variables for any given time ordering. The resulting expression is plugged back into the integration over all dummy variables $t_m$. Since we have a time-ordered (and anti-time ordered) set of integrals the integration limits need to be adjusted accordingly. The result is,

\begin{align}
A^{(n)} = n \int_{-t_0}^t dt' \sum_{m=0}^{n-1}(-1)^{m} \binom{n-1}{m} \int_{-t_0}^{t'}\dint{n} \Bigg\lbrace~ \nonumber\\
\bar{\mathcal{T}}\Bigl[\prod_{j=1}^{m} \htu(t_j)\Bigr] 
\bigl[\mo(t),\htu(t')\bigr]\mathcal{T}\Bigl[\prod_{j=m+1}^{n-1} \htu(t_j)\Bigr]\Bigg\rbrace\label{eqn:ordering}
\end{align}

The upper limit of the integration variables $t_m$ is $t'$, which is the label of $\htu$ appearing in the commutator. An extra factor of $n$ appears because we are summing over all possible (anti-)time orderings. Plugging this expression back into the original expansion \eqref{eqn:expansion} results in
\begin{align}
\sum_{n=1}^\infty \frac{(-i)^n}{n!} A^{(n)} = 
-i\int_{t_0}^t \sum_{n=0}^\infty \frac{(-i)^{n}}{n!}  B^{(n)}(t')~ dt' \label{app:almost_final_result}
\end{align}
where $B^{(0)}(t') = \left[\mo(t),\htu(t')\right]$ and for $n>0$
\begin{multline}
B^{(n)}(t') = \sum_{m=0}^{n}(-1)^{m} \binom{n}{m} \int_{-t_0}^{t'}\dint{n}\bar{\mathcal{T}}\Bigl[\prod_{j=1}^{m} \htu(t_j)\Bigr]\\
\times \left[\mo(t),\htu(t')\right]\mathcal{T}\Bigl[\prod_{j=m+1}^{n-1} \htu(t_j)\Bigr]~.
\end{multline}
The summation over $B^{(n)}$ matches that of Eq.~\eqref{eqn:expansion}, but with $\mo(t)$ replaced by $ \left[\mo(t),\htu(t')\right]$. The right hand side of Eq.~\eqref{app:almost_final_result} is therefore equal to
\begin{multline}
-i\int_{t_0}^t  \sum_{n=0}^\infty \frac{(-i)^{n}}{n!}  B^{(n)}(t')~dt' = \\
-i\int_{t_0}^t  \mathcal{U}^\dag(t,t_0)[\mathcal{O}(t),\htu(t')]\mathcal{U}(t,t_0)~dt'
\end{multline}
This proves the non-equilibrium Kubo formula \eqref{app:gen-result}.

\section{NON-EQUILIBRIUM WARD IDENTITY AND CROSS CORRELATIONS}\label{app2:crosscorr}
In this appendix we determine an expression for the correlator $\Delta S$, which appears in the finite frequency noise of the edge current, see equation \eqref{eq:mainresult1}. This is done by making use of a Ward identity for the correlators involved.

\subsection{Cross correlation}
The correlator $\Delta S$ appears in Eq.~\eqref{crossnoise} and is given by the Fourier transform of
\begin{align}
\Delta S(t) &= \mathcal{F}(t) + \mathcal{F}(-t-i/T) \nonumber\\ 
\mathcal{F}(t) &= \langle \Delta j_R(x,t+x/v_c)\ib(0)\rangle \label{app:deltast}
\end{align}
Here we made use of the Kubo-Martin-Schwinger relation\citep{rammer2007} which relates $\langle I_B^I(0)  \Delta j_R(x,t)\rangle_0 = \langle \Delta j_R(x,t-i/T) I_B^I(0)\rangle_0$.  The KMS relations is explained in Section~\ref{subsection_noneqfdt}, see Eq.~\eqref{eq:kms}. It is a condition on two-point correlators in equilibrium systems. Using the KMS relation we can write
\begin{align}
\Delta S(\omega) = \mathcal{F}(\omega) + e^{\omega/T} \mathcal{F}(-\omega)\label{app:deltas}
\end{align}
Making the time evolution operators $\ut$ explicit we have for $\mathcal{F}(\omega)$,
\begin{align}
\mathcal{F}(t) &=  \langle \Delta j_R(t) \ut^\dag(0,-\infty) \hat{I}_B(0) \ut(0,-\infty)\rangle \nonumber\\
\mathcal{F}(\omega) &= \int e^{i\omega t} \mathcal{F}(t)   dt\label{app:s1}
\end{align}
Here $\Delta j_R(t) = j_R(x,t+x/v_c) - \langle j_R\rangle = -v_c\frac{\snu}{2\pi}\partial_x\ph_R$, and we have suppressed the spatial dependence as it drops out in the end. Our goal is to simplify the expression \eqref{app:s1}. This accomplished by making use of the Ward identity associated with the current $j$.

\subsection{The equilibrium Ward identity}

Ward identities are restrictions imposed on correlation functions in a theory as a consequence of symmetries of the theory. In our case the insert operator $\Delta j_R$ is a conserved current associated with the $u(1)$ symmetry. In complex coordinates on the plane the Ward identity is given by\citep{knizhnik1984}
\begin{multline}
\langle \partial_z\ph(z) \psi_1(z_1)\cdots \psi_n (z_n) \rangle = \\ \sum_{j} \frac{\tilde{Q}_j}{z-z_j} \langle  \psi_1(z_1)\cdots \psi_n(z_n)  \rangle ~.\label{app:complexward}
\end{multline}
Each operator $\psi_i$ contributes a term $\frac{\tilde{Q}_j}{z-z_j}$ times the correlator without the current. We call this a contraction. Here $\tilde{Q}_j$ is the charge of the operator with respect to the current $\partial_z\ph$. In our case it equals the electric charge up to normalization of the current operator.

To obtain a finite-temperature correlator we perform a conformal mapping to the cylinder, $z=\exp(2\pi T i w/v_c)$, on the left and right-hand side of Eq.~\eqref{app:complexward}. Each field transforms covariantly as\citep{belavin1984,difrancesco1995}
\begin{align}
\psi_j(z_j) = \Bigl(\frac{dw_j}{dz_j}\Bigr)^{h_j}\psi_j(w_j)\label{eq:covariant_transform}
\end{align}
with $h_j$ the conformal dimension of the operator. But since the $\psi_j$ operators appear on both the left- and right-hand side all factors of $(dz_j/dw_j)^{h_j}$ cancel, except for one factor of $dz/dw$ which is associated with the current. The current has $h=1$ and so the Ward identity on the cylinder is
\begin{multline}
\langle \partial_w\ph(w) \psi_1(w_1)\cdots \psi_n (w_n) \rangle = \\ 
%\sum_{j} \frac{\tilde{Q}_j\frac{dz}{dw}}{e^{2\pi T i w/v_c}-e^{2\pi T i w_j/v_c}} \langle  \psi_1(w_1)\cdots \psi_n(w_n)  \rangle ~ \\
%\sum_{j} \frac{ 2\pi T i\tilde{Q}_j /v_c \exp(2\pi T i w/v_c)}{e^{2\pi T i w/v_c}-e^{2\pi T i w_j/v_c}} \langle  \psi_1(w_1)\cdots \psi_n(w_n)  \rangle ~ \\
\sum_{j} \frac{iT \tilde{Q}_j /v_c}{1-e^{-2\pi T i (w - w_j)/v_c}} \langle  \psi_1(w_1)\cdots \psi_n(w_n)  \rangle ~.
\label{app_ward_almost}
\end{multline} 
The final step is to transform this equation to real time. For this we replace \mbox{$w-w_j$} by \mbox{$\delta +i v_c(t-t_j)$}, where $\delta$ is an integral regulator (we set $x_j = 0$ for all $j$). We also identify \mbox{$v_c\frac{\snu}{2\pi }\partial_w \ph(w) \leftrightarrow  \Delta j_R(x, t-x/v_c)$}. The charge $\tilde{Q}_j$ relates to the electric charge as $\frac{\snu}{2\pi}\tilde{Q}_j = Q_j$. When everything is put together we obtain
\begin{multline}
\langle \Delta j_R(t) \psi_1(t_1)\cdots \psi_n(t_n)\rangle = \\
\sum_j Q_j K(t-t_j) \langle  \psi_1(t_1)\cdots \psi_n(t_n)\rangle \label{eq_ward}
\end{multline}
where $K$ follows from Eq.~\eqref{app_ward_almost}. However, $K$ is not uniquely determined because of the neutrality condition which states $\sum_j Q_j = 0$. We can add any constant to $K$ and Eq.~\eqref{eq_ward} remains valid. We set
\begin{align}
K(t) &=
% \left.\frac{\bigl(\frac{d}{dw}e^{2\pi T i w/v_c}\bigr)}{e^{2\pi T i w/v_c}-e^{2\pi T i w_i/v_c}}
%\right|_{\frac{w-w_i}{v_c} = \delta + i(t-t_i)}\nonumber\\
 %\frac{i T}{1-e^{2\pi T (t-i\delta)}}
\frac{T}{2} \cot\bigl(\pi T(\delta + i t)\bigr)\label{real_time_ward}
\end{align}
In the limit of $\delta \downarrow 0$ this expression is antisymmetric. This is the real-time, finite temperature version of the Ward identity Eq.~\eqref{app:complexward}. If there are multiple channels then we have a Ward identity for each channel, with the charge appropriately weighted. Also, any neutral part of the operator $\psi_i$ is carried along without affecting the end result, so the result applies to non-Abelian states as well.

\subsection{Non-equilibrium Ward identity}
The Ward identity \eqref{real_time_ward} applies to correlators in which the time evolution of the operators is due to the equilibrium Hamiltonian. The operators that enter the expression of $\mathcal{F}$, see Eq.~\eqref{app:s1}, are in the interaction representation. We therefore need to extend the Ward identity to this interaction picture. To accomplish this we expand the correlators using the series expansion of the time evolution operators $\mathcal{U}$ and apply the Ward identity term-by-term.

Both $\htu$ and $\ib$ are given in terms of $\V$ and $\V^\dag$. Furthermore, $\V \propto \psi_L^\dag\psi_R$ and so the operator $\V$ ($\V^\dag$) carries a charge of $-Q$ ($Q$) with respect to $j_R$. Therefore, whenever the correlator contains a tunnelling Hamiltonian $\htu$ we have
\begin{multline}
\langle \Delta j_R(t) \cdots \htu(t')\cdots \rangle = \\
 i K(t-t') \langle \cdots\hat{I}_B(t') \cdots \rangle + \ldots  \label{app:contraction_with_htu}
\end{multline}
%For the multichannel case this relation reads
%\begin{multline}
%\langle \Delta j_R(t) \cdots \htu(t')\cdots \rangle = \\
% i K(t-t') \frac{\eta_i \kappa_i q_i}{Q}\langle \cdots\hat{I}_B(t') \cdots \rangle + \ldots  
%\end{multline}
Here the dots represent the remaining contractions. A similar expression holds for the tunnelling operator $\hat{I}_B(t)$ in which case $\ib$ is replaced by $-iQ^2\htu(t)$. We now apply this result to \eqref{app:s1}. First we expand the operators $\mathcal{U}$ and $\mathcal{U}^\dag$. This results in
\begin{multline}
\mathcal{F}(t) = 
\sum_{n,m=0}^\infty \frac{(i)^n}{n!}\frac{(-i)^m}{m!} \Bigl[\prod_{i=1}^n\int_{-\infty}^0  dt_i \Bigr]
\Bigl[\prod_{j=1}^m\int_{-\infty}^0  dt_j' \Bigr] \\
\langle \Delta j(t) \ov{\mathcal{T}}\Bigl[\prod_{i=1}^n\htu(t_i)\Bigr] \hat{I}_B(0)\mathcal{T}\Bigl[\prod_{j=1}^m\htu(t_j')\Bigr] \rangle \label{app:expansion_of_f}
\end{multline}
with $\mathcal{T}[\cdot]$ and $\ov{\mathcal{T}}[\cdot]$ time and reversed-time ordering operators. Applying the Ward identity results in
\begin{widetext}
\begin{align}
&\langle \Delta j(t) \ov{\mathcal{T}}\Bigl[\prod_{i=1}^n\htu(t_i)\Bigr] \hat{I}_B(0)\mathcal{T}\Bigl[\prod_{j=1}^m\htu(t_j')\Bigr] \rangle = -iQ^2 K(t)\langle  \ov{\mathcal{T}}\Bigl[\prod_{i=1}^n\htu(t_i)\Bigr] \htu(0)\mathcal{T}\Bigl[\prod_{j=1}^m\htu(t_j')\Bigr] \rangle \label{wide_text_equation}\\
&i\sum_{k=1}^n 
K(t-t_k)\langle \ov{\mathcal{T}}\Bigl[I_B(t_k)\prod_{\substack{i=1\\ i\neq k}}^n\htu(t_i)\Bigr] \hat{I}_B(0)\mathcal{T}\Bigl[\prod_{j=1}^m\htu(t_j')\Bigr] \rangle +
 i\sum_{k=1}^m K(t-t_k')
 \langle  \ov{\mathcal{T}}\Bigl[\prod_{i=1}^n\htu(t_i)\Bigr] \hat{I}_B(0)\mathcal{T}\Bigl[\hat{I}_B(t_k')\prod_{\substack{j=1\\ j\neq k}}^m \htu(t_j')\Bigr] \rangle  \nonumber
\end{align}
\end{widetext}
The first term comes from the contraction of $j_R$ with $\ib(0)$. The other two terms are the contractions of $j_R$ with the $\htu$ appearing in the time evolution operators. We plug the total expression Eq.~\eqref{wide_text_equation} back into the summations and integrations in Eq.~\eqref{app:expansion_of_f}. Our next goal is to show that this step results in the following non-equilibrium Ward identity
\begin{multline}
\mathcal{F}(t) = -i Q^2 K(t) \langle \htu^I(0)\rangle \\
-\int_{-\infty}^0  K(t-t')\langle [\ib(t'), \ib(0)] \rangle ~dt'~.\label{non_eq_ward}
\end{multline}
We are interested in the summation and integration over Eq.~\eqref{wide_text_equation}, i.e. 
\begin{align}
\sum_{n,m=0}^\infty \frac{(i)^n}{n!}\frac{(-i)^m}{m!} \Bigl[\prod_{i=1}^n\int_{-\infty}^0  dt_i \Bigr]
\Bigl[\prod_{j=1}^m\int_{-\infty}^0  dt_j' \Bigr] \Big[\text{Eq.~\eqref{wide_text_equation}}\Big]\nonumber
\end{align}
Consider the first term appearing in Eq.~\eqref{wide_text_equation} (proportional to $Q^2$). It should be straightforward to see that this term results in the first term of Eq.~\eqref{non_eq_ward}. Next we consider the second term of Eq.~\eqref{wide_text_equation}. The integration over $dt_j'$ and summation over $m$ results in $\mathcal{U}(0,-\infty)$. What remains is
\begin{multline}
\sum_{n=1}^\infty \frac{(i)^{n-1}}{n!} \Bigl[\prod_{i=1}^n\int_{-\infty}^0  dt_i \Bigr] 
\sum_{k=1}^n K(t-t_k)\\
\times\langle \ov{\mathcal{T}}\Bigl[I_B(t_k) \prod_{\substack{i=1\\ i\neq k}}^n\htu(t_i)\Bigr] \hat{I}_B(0)\mathcal{U}(0,-\infty) \rangle \label{what_remains}
\end{multline}
By changing integration variables ($t_k\rightarrow t'$ and some additional relabelling) we can write this expression as
\begin{multline}
-\int_{-\infty}^0  dt'K(t-t')   \sum_{n=1}^\infty \frac{(i)^{n-1}}{(n-1)!} \Bigl[\prod_{i=1}^{n-1}\int_{-\infty}^0  dt_i \Bigr]
\\
 \langle \ov{\mathcal{T}}\Bigl[I_B(t') \prod_{i=1}^{n-1}\htu(t_i)\Bigr] \hat{I}_B(0)\mathcal{U}(0,-\infty) \rangle
\end{multline}
The final integration and summation results in
\begin{multline}
 \sum_{n=0}^\infty \frac{(i)^{n}}{n!} \prod_{i=1}^{n}\int_{-\infty}^0  dt_i \ov{\mathcal{T}}\Bigl[I_B(t') \prod_{i=1}^{n}\htu(t_i)\Bigr] =\\ \ov{\mathcal{T}}\left[\hat{I}_B(t')  \mathcal{U}^\dag(0,-\infty)\right] = \mathcal{U}^\dag(t',-\infty) \hat{I}_B(t') \mathcal{U}^\dag(0,t')  ~.
\end{multline}
Finally, combining this result with Eq.~\eqref{what_remains} results in
\begin{align}
[\text{2$^{\text{nd}}$ term}] \longrightarrow -\int_{-\infty}^0  dt'K(t-t')
\langle \ib(t')  \ib(0) \rangle
\end{align}
The manipulation of the third and final term in Eq.~\eqref{wide_text_equation} is done along the same lines and results in
\begin{align}
[\text{3$^{\text{rd}}$ term}] \longrightarrow \int_{-\infty}^0  dt'K(t-t')\langle \ib(0) \ib(t')  \rangle
\end{align}
Putting everything together results in the non-equilibrium Ward identity Eq.~\eqref{non_eq_ward}.

Next we look at the Fourier transform of $\mathcal{F}(t)$. To obtain this we require $K(\omega)$. This can be obtained for instance through a contour integral. The result is
\begin{align}
\int e^{i\omega t} K(t) dt
%& = -i\frac{T}{2} \int e^{i\omega t}  
%\frac{1+e^{-2\pi T i(\delta + it)}}{1-e^{-2\pi T i(\delta + it)}} dt  \nonumber\\
%& = -i\frac{1}{4\pi} \int e^{i\frac{\omega}{2\pi T} x}  
%\frac{1+e^{x-2\pi T i\delta }}{1-e^{x-2\pi T i\delta }} dx  \nonumber\\
%& = 
%-
%\theta(\omega)
%e^{-\omega\delta} \sum_{n=0}^{\infty} e^{-\frac{\omega}{T} n} \\
%&+
%\theta(-\omega)
%e^{-\omega\delta} \sum_{n=1}^{\infty} e^{\frac{\omega}{T} n}
%\nonumber\\
%&=
%-\theta(\omega)e^{-\omega\delta} \frac{e^{\frac{\omega}{T}}}{e^{\frac{\omega}{T}}-1}\nonumber\\
%&-\theta(-\omega)e^{-\omega\delta} \frac{e^{\frac{\omega}{T}}}{e^{\frac{\omega}{T}}-1}
%\nonumber\\
&= - e^{-\omega\delta} \frac{e^{\frac{\omega}{T}}}{e^{\frac{\omega}{T}}-1}
\nonumber\\
&= -\frac{1}{2}   e^{-\delta\omega}\bigl( \coth\bigl(\frac{\omega}{2T}\bigr) +1\bigr) \nonumber\\
&\equiv -\frac{1}{2} e^{-\delta \omega} N(\omega)~.
\end{align}
The frequency representation of the non-equilibrium Ward identity is then (taking $\delta \downarrow 0$)
\begin{align}
\mathcal{F}&(\omega) = 
\frac{1}{2}  N(\omega) \label{app:f_omega}\\
&\times
\Bigl[i Q^2 \langle \htu^I(0)\rangle + \int_{-\infty}^0 e^{i\omega t'} 
\langle [\ib(t'), \ib(0)] \rangle   dt' \Bigr]~. \nonumber
\end{align}
Note also the appearance of the antisymmetric noise, $R_{I_B}$, in the expression for $\mathcal{F}$. For $\Delta S$ we use Eq.~\eqref{app:deltas} and obtain  
%(1+\coth(x)) = -(1+coth(-x))e^{2x}
\begin{align}
\Delta S(\omega) = 
N(\omega) R_{I_B}(\omega)~. \label{app:deltas_relation}
\end{align}
This proves the relation \eqref{crossnoise}. We also note the real and imaginary parts of $\mathcal{F}$
\begin{align}
2\text{Re}\bigl[\mathcal{F}(\omega)\bigr] &= \Delta S(\omega)\\
2\text{Im}\bigl[\mathcal{F}(\omega)\bigr] &= Q^2N(\omega)\langle \htu^I(0)\rangle~.
\end{align}
and so $\mathcal{F}(\omega)^* = e^{\omega/T} \mathcal{F}(\omega)$. 

\subsection{Multichannel case}
We also comment on the multichannel case. In this case we have a non-equilibrium Ward identity for each channel. The difference is that the current operator of the $i$'th channel only measures a fraction of the total charge of the tunnelling operator $\V$. In particular, expression \eqref{app:contraction_with_htu} becomes
\begin{multline}
\langle \Delta j_i(t) \cdots \htu(t')\cdots \rangle = \\
 i \frac{\kappa_i q_i}{Q} K(t-t') \langle \cdots\hat{I}_B(t') \cdots \rangle + \ldots  
\end{multline}
The final identity Eq.~\eqref{app:f_omega} is scaled down by the same factor of $\kappa_i q_i/Q$. In the treatment of the multichannel case we also encounter the following cross correlation which mixes velocities of different channels
\begin{multline}
\Delta S_{ij}(t)  = \langle\Delta j_{i}(x,t+\eta_j x/ v_j) \ib(0)  \rangle \\  
+
\langle  \ib(0) \Delta j_{i}(x,-(t+\eta_j x/ v_j)) \rangle  ~.
\end{multline}
This requires a bit more care, as we encounter the velocity $v_j$ instead of $v_i$ (compare this to Eq.~\eqref{app:deltast}). Using the KMS relation Eq.~\eqref{eq:kms} we obtain 
\begin{multline}
\Delta S_{ij}(t) = \frac{\kappa_i q_i}{Q}\mathcal{F}\Bigl(t-x\bigl(\frac{\eta_i }{ v_i } -\frac{\eta_j }{ v_j}\bigr) \Bigr) \\+
\frac{\kappa_i q_i}{Q}\mathcal{F}\Bigl(-t-x\bigl(\frac{\eta_i }{ v_i } -\frac{\eta_j }{ v_j}\bigr) - i/T \Bigr)
\end{multline}
and its Fourier transform
\begin{align}
\Delta S_{ij}(\omega)  
%= 
% \frac{\kappa_i q_i}{Q}
% e^{i\omega x\bigl(\frac{\eta_i }{ v_i } -\frac{\eta_j }{ v_j}\bigr) } (\mathcal{F}(\omega) + e^{\omega/T}\mathcal{F}(-\omega)) \\
 =  \frac{\kappa_i q_i}{Q}
 e^{i\omega x\bigl(\frac{\eta_i }{ v_i } -\frac{\eta_j }{ v_j}\bigr) } N(\omega) R_{I_B}(\omega)~.
\end{align}

\bibliography{./mainbib}

\end{document}